\documentclass[review]{elsarticle}

\usepackage[utf8]{inputenc}
\usepackage[T1]{fontenc}
\usepackage[a4paper,top=3cm,bottom=2cm,left=3cm,right=3cm,marginparwidth=1.75cm]{geometry}
\usepackage{amssymb}
\usepackage{amsmath}
\usepackage{graphicx}
\usepackage[colorlinks=true, allcolors=blue]{hyperref}

\usepackage{caption}
\usepackage{subcaption}
\usepackage{pgfplots}
\pgfplotsset{compat=1.14}

\emergencystretch=\maxdimen
\begin{document}

\begin{frontmatter}

\title{The Forgotten Document-Oriented Database Management Systems:\\ An Overview and Benchmark of Native XML DODBMSes in Comparison with JSON DODBMSes}

\author[1]{Ciprian-Octavian~Truică\corref{c1}\fnref{c2,c3}}

\cortext[c1]{Corresponding author.}
\fntext[c2]{These authors contributed equally to this article.}
\fntext[c3]{Part of this work was done at Aarhus University.}
\ead{ciprian.truica@cs.pub.ro}

\author[1]{Elena-Simona~Apostol\corref{c1}\fnref{c2}} 
\ead{elena.apostol@cs.pub.ro}

\address[1]{Computer Science and Engineering Department, Faculty of Automatic Control and Computers, University Politehnica of Bucharest, Romania}
\address[2]{Department of Computer Science, Aarhus University, Aarhus, Denmark}

\author[3]{Jérôme~Darmont}
\address[3]{Université de Lyon, Lyon 2, ERIC UR 3083, France}
\ead{jerome.darmont@univ-lyon2.fr}

\author[4]{Torben~Bach~Pedersen}
\address[4]{Center for Data Intensive Systems, Aalborg University, Aalborg, Denmark}
\ead{tbp@cs.aau.dk}

\begin{abstract}
   In the current context of Big Data, a multitude of new NoSQL solutions for storing, managing, and extracting information and patterns from semi-structured data have been proposed and implemented.
   These solutions were developed to relieve the issue of rigid data structures present in relational databases, by introducing semi-structured and flexible schema design.
   As current data generated by different sources and devices, especially from IoT sensors and actuators, use either XML or JSON format, depending on the application, database technologies that store and query semi-structured data in XML format are needed.
   Thus, Native XML Databases, which were initially designed to manipulate XML data using standardized querying languages, i.e., XQuery and XPath, were rebranded as NoSQL Document-Oriented Databases Systems.
   Currently, the majority of these solutions have been replaced with the more modern JSON based Database Management Systems.
   However, we believe that XML-based solutions can still deliver performance in executing complex queries on heterogeneous collections.
   Unfortunately nowadays, research lacks a clear comparison of the scalability and performance for database technologies that store and query documents in XML versus the more modern JSON format.
   Moreover, to the best of our knowledge, there are no Big Data-compliant benchmarks for such database technologies.
   In this paper, we present a comparison for selected Document-Oriented Database Systems that either use the XML format to encode documents, i.e., BaseX, eXist-db, and Sedna, or the JSON format, i.e., MongoDB, CouchDB, and Couchbase.
   To underline the performance differences we also propose a benchmark that uses a heterogeneous complex schema on a large DBLP corpus.
\end{abstract}

\begin{keyword}
    XML Database Management Systems; JSON Database Management Systems; Document-Oriented Database Management Systems; Benchmark
\end{keyword}

\end{frontmatter}

\section{Introduction}\label{sec:intro}

With the emergence of Big Data and the Internet of Things (IoT) and the increasing amount of semi-structured information generated daily, new technologies have arisen for storing, managing, and extracting information and patterns from such data.
The new technologies for storing data have been labeled with the name NoSQL and were initially developed to solve very specific problems.
Currently, they provide different trade-offs and functionality (e.g., choosing high-availability over consistency) to be as generic as their counterparts Relational Database Management Systems (RDBMSes).
Due to the semi-structured nature of data, NoSQL Database Management Systems (DBMSes) have been classified based on the data model used for storing information~\cite{Han2011}, i.e., key-value, document-oriented, wide column, and graph databases.

In this paper, we particularly study NoSQL Document-Oriented Databases Systems (DODBMSes) that encode data using the XML or JSON formats.
We further focus on two subcategories of DODBMSes with respect to the data model used to encode documents: 
i) DODBMSes that encode data using the XML format are Native XML Database Management Systems (XDBMSes), and
ii) DODBMSes that encode data using the JSON format are JSON Database Management Systems (JDBMSes).

The NoSQL DBMSes became very popular with the increasing need for data storage, management, and analysis systems that scale with the volume.
To address these needs, many NoSQL DBMSes compromise consistency to offer high-availability, partition tolerance, improved analytics, and high-throughput.
These features are also a requirement for real-time web applications and Big Data processing and analysis and are available in JDBMSes as well.

XDBMSes have started to emerge after the eXtensible Markup Language (XML) has been standardized and became the common format for exchanging data between different applications running on the Web.
Their primary use was to facilitate secure storage and fast querying of XML documents.
Besides their primary use, XDBMSes prove useful for OLAP (Online Analytical Processing) style analysis and decision support systems that incorporate a time dimension and encode data in the XML format~\cite{Park2005}, and thus removing the need of using ETL (Extract Transform Load) processes to transform XML documents into a relational model.
XML query languages and technologies, including XDBMSes, had been around before the NoSQL trend, and have been forgotten during the Big Data hype.
In the field of relational databases, XML format is used as a Data Type, e.g., Oracle, DB2, PostgreSQL, etc.
Currently, with the rise of the NoSQL movement, XDBMSes have become a subcategory of DODBMSes.
But, with the emergence of processing platforms that uses Big Data or IoT technologies, where the data are transferred over computer networks into formats such as XML and JSON, the XDBMSes can be seen as a viable solution for storing and manipulating computer-generated semi-structured data.

We hypothesize that the more classical XDBMSes may still be useful in the Big Data era.
Thus, in this study we want to address and use as guidelines the following research questions:
\begin{itemize}
    \item [\textbf{Q1:}] Are XDBMSes absolute and should be replaced by JDBMSes?
    \item [\textbf{Q2:}] Are XDBMSes a viable candidate for Big Date Management?
    \item [\textbf{Q3:}] Do JDBMSes outperform XDBMSes when using complex filtering and aggregation queries with different scale factors, on large and heterogeneous datasets?
\end{itemize}

To test our hypothesis and answer our research questions, we consider the following research objectives:
i) discuss XDBMSes and compare their capabilities and features with several popular JDBMSes solutions;
ii) propose a benchmark that evaluates the current needs and workloads available in Big Data and compare performance between the selected DODBMSes;
iii) evaluate the performance of the selected DODBMSes using complex filtering and aggregation queries with different scale factors, on large and heterogeneous datasets.

For testing and analyzing with our proposed benchmark, we utilize several XDBMSes and JDBMSes solutions, that are free to use, and their license does not forbid benchmarking.
Thus, we chose BaseX, eXist-db, and Sedna as representatives XDBMSes systems and MongoDB, CouchDB, and Couchbase as JDBMSes solutions.

As a result of our research and as a response to \textbf{Q1}, we claim that the more classical XML based DODBMSes may still be useful in the Big Data era.
To demonstrate this and answer \textbf{Q2}, we propose a new benchmark for comprehensive DODBMSes analysis using a large dataset .
And thereby we present a qualitative and quantitative performance comparison between XDBMSes and the more modern JDBMSes to answer \textbf{Q3}.

This paper is structured as follows.
Section~\ref{sec:sota} presents an overview of different NoSQL DBMSes models, surveys, and benchmarks.
Section~\ref{sec:comparison} offers an in-depth overview and comparison of DODBMSes, focusing on the XDBMSes and JDBMSes subcategories.
Section~\ref{sec:specs} introduces the proposed benchmark specification and discusses the data and workload models, while Section~\ref{sec:impl} discusses the database physical implementation and presents the description of the benchmark's queries.
Section~\ref{sec:experiments} thoroughly details the experiments performed on the selected DODBMSes using our benchmark and discusses the results in detail.
Finally, Section~\ref{sec:conclusion} concludes the paper, summarizes the results, and provides future research perspectives.

\section{Related Works}\label{sec:sota}

The NoSQL Database Management Systems (DBMSes) emerged as an alternative to Relational Database Management Systems (RDBMSes) in order to store and process huge amounts of heterogeneous data.
However, NoSQL DBMSes did not appear as a replacement for RDBMSes, but as a solution to specific problems that require additional features (e.g., replication, high-availability, etc.) that are not handled well by traditional means~\cite{Stonebraker2005}.
The reasons commonly given to develop and use NoSQL DBMSes are summarized as follows~\cite{Strauch2011}: avoidance of unneeded complexity, high throughput, horizontal scalability, running on commodity hardware, avoidance of expensive object-relational mapping, lowering the complexity and the cost of setting up a cluster, compromising reliability for better performance, and adapting to the requirements of cloud computing.

The classifications used for NoSQL DBMSes usually are done by either taking into account the persistence model or the data and query model.
Using the persistence model, NoSQL DBMSes are classified as follows~\cite{Strauch2011}: 
\begin{itemize}
\item[i)] 
    In-Memory Databases~\cite{Zhu2018} are very fast because the most current used data are stored in memory, with optional subsequent disk flushes triggered at given periods or when the in-memory data are not used.
    Evidently, the size of the currently in-use data that can be stored is limited to the amount of memory.
    This problem can be resolved using vertical scaling to some degree as there is a limit to the amount of memory a system can hold.
    Moreover, the durability may become a problem if data are lost between subsequent disk flushes or if data persistence is disabled.
    A solution to this problem is data replication.
\item[ii)] 
    Memtables and SSTables Databases~\cite{Qader2018} buffer operations in memory using a Memtable after they have been written to an append-only commit log to ensure durability.
    After a certain amount of writes the Memtable gets flushed to disk as a whole into a SSTable.
    These DBMSes have performance characteristics comparable to those of In-Memory Database but solve the durability problem.
\item[iii)] 
    B-trees Databases~\cite{Petrov2018} use the B-tree self-balancing tree data structure that keeps data sorted and allows searches, sequential access, insertions, and deletions in logarithmic time~\cite{Comer1979}.
\end{itemize}

NoSQL DBMSes are also classified by using the data and query model as follows~\cite{Han2011,Cattell2011}:
\begin{itemize}
\item[i)] 
    Wide Column Databases are used to store, retrieve, and manage data using column families.
    Each record can have different numbers of cells and columns, making a row sparse without storing NULLs.
\item[ii)] 
    Graph Databases are used to store, retrieve, and manage information using a graph.
    Therefore, an object is modeled as a node and the edges between nodes become the relationships between the objects.
\item[iii)] 
    Key-Value Databases (KVDBMSes) are data storage systems designed for storing, retrieving, and managing associative arrays, i.e., dictionaries or hash tables.
\item[iv)] 
    Document-Oriented Databases (DODBMSes) have evolved form KVDBMSes and are used to store, retrieve, and manage semi-structured data, i.e., documents, encoded using JSON, BSON, XML, or YAML formats.
\end{itemize}

There are multiple surveys on NoSQL DBMSes, in the following phrases we present the most relevant ones for our analysis.
Article~\cite{Stonebraker2010} provides a comparison regarding the performance and flexibility of KVDBMSes and DODBMSes over RDBMSes.
The NoSQL DBMSes prove to be a better choice for high throughput applications that require data modeling flexibility and horizontal scaling.
The authors of~\cite{Han2011} offer a classification by data models of NoSQL DBMSes, and also they present the current and most popular solutions.
In~\cite{Hecht2011}, the authors make a comparison and overview of NoSQL data models, query types, concurrency controls, partitioning, and replication.
Article~\cite{Gessert2016} presents a top-down overview of the NoSQL database field and propose a comparative classification model that relates functional and non-functional requirements to techniques and algorithms employed in these systems.
The authors of~\cite{Brahmia2020} provide an overview of XML data manipulation techniques employed in conventional and temporal XDBMSes and study the support of such functionality in mainstream commercial DBMSes.
Unfortunately, the paper presents only a general discussion about XDBMSes and other DBMSes with XML manipulation capabilities, and also no evaluation is provided.
Thus, we can conclude that none of these surveys present an in-depth discussion and comparison of different subcategories of DODBMSes.

In the literature there are many data-centric benchmarks for the Big Data distributed systems and NoSQL DBMSes that focus either on structured data or on specific applications, such as MapReduce-based applications, rather than on unstructured or variety.
In~\cite{Bajaber2020}, the authors present a comprehensive survey and analysis of benchmarks for different types of Big Data systems including NoSQL systems.
The authors of \cite{Truica2020} present a new benchmark for textual data for distributed systems including MongoDB.
None of the current literature presents benchmarks for modern native XDBMSes.

XDBMSes benchmarks are application-oriented and domain-specific, e.g.,
OpenEHR XML medical records~\cite{Freire2012}, 
XMark which contains documents extracted from electronic commerce sites and content providers~\cite{Schmidt2002}
or Transaction Processing over XML (TPoX)~\cite{Nicola2007} which simulates a financial multi-user workload with XML data conforming to the FIXML standard.
These benchmarks are used for testing the performance of DBMSes that are capable of storing, searching, modifying and retrieving XML data.
Unfortunately, the majority of these benchmarks use rather small collections.
And even for the benchmarks where the XML or JSON document size is up to the order of Gigabytes (GBs), the contained information is mostly homogeneous.
Our proposed benchmark solution uses large heterogeneous collections with over 6 million records to test the scalability, filtering, and aggregation performance of complex queries for the current native XDBMSes.

Based on the lack of current literature regarding XDBMS, in this paper, we analyze the performance and functionality of DODBMSes solutions, while focusing on two distinct subclasses that use JSON or XML formats to encode data.

\section{Document-Oriented Databases}\label{sec:comparison}

Document-Oriented Databases Management Systems (DODBMSes) have evolved from Key-Value Databases~\cite{Han2011}.
DODBMSes are used for storing, retrieving, and managing semi-structured data.
They have a schema-less flexible data representation, thus providing more flexibility for data modeling~\cite{Atzeni2020}.
DODBMSes use documents for storing data such as XML or JSON.
The flexibility provided by XML and JSON makes it easier to manipulate the information than it is for tables in Relational Database Management Systems (RDBMSes).
Usually, documents are stored in collections.
A Native XML Database Management System (XDBMS) uses the XML (eXtensible Markup Language) data structure to encode documents and defines a hierarchical logical model based on the elements of this markup language~\cite{Fiebig2002,Pavlovic2007}.
A JSON Database Management System (JDBMS) uses the JSON structure for modeling documents and storing them in collections.

In DODBMSes, labels are used in storing the information.
These labels describe the data and values in a record.
New information can be added directly to a record without the need to modify the entire schema, as is the case for RDBMSes.

One of the benefits of using a DODBMS solution is the flexibility of modeling the data~\cite{Gallinucci2018}.
Data from the web, mobile, social, and IoT devices change the nature of the application's data model.
In an RDBMS, these changes impose the modification of the schema by altering tables and adding or removing columns.
Whereas, the flexibility of DODBMSes eliminates the need to force-fit the data into predefined attributes and tables.

Another benefit of a DODBMS is the fast write performance.
Some DODBMSes prioritize high availability over strict data consistency.
This ensures that both read and write operations will always be executed even if there is a hardware or network failure.
In case of failure, the replication and eventual consistency mechanisms ensure that the environment will function.

Fast query performance is another benefit of a DODBMS.
Most DODBMSes provide powerful query engines for CRUD (Create, Read, Update and Delete) operations and use indices and secondary indices to improve data retrieval.
Additionally, the majority of DODBMS solutions support aggregation frameworks, either native or using MapReduce, for Data Analysis and Business Intelligence.

\subsection{XDBMSes}

In this subsection, we present several examples of XDBMSes that use standardized XPath and XQuery.
Although there are multiple solutions of DBMSes that incorporate XML as data type (e.g., Oracle, PostgreSQL, DB2, MS SQL, etc. just to name a few), the majority of them fall out of the NoSQL movement.
Furthermore, some have licenses that explicitly forbids benchmarking, e.g., commercial XDBMSes such as MarkLogic Server and Oracle Berkeley DB XML.
Thus, for our comparison and benchmark, we chose the following three XDBMSes: BaseX, eXist-db, and Sedna.

\subsubsection*{BaseX}

BaseX is an XDBMS written in Java that stores the data using a schema-free hierarchical model.
Transactions in BaseX respect the ACID (Atomicity, Consistency, Isolation, and Durability) properties, enabling the concurrent access of multiple readers and writers~\cite{BaseX}.
Documents are stored either persistently on disk or in the main memory.
BaseX uses a single instance environment, replication and data partitioning are not available.

BaseX provides CRUD operations and ad-hoc queries, including aggregation using XQuery 3.1 and XPath 3.1~\cite{Grun2009}.
Although, it works with various APIs such as XML DB or JAX-RX, it was not designed to work with a MapReduce framework.
 
BaseX supports multiple structural and value indices~\cite{BaseX}.
Structural indices are automatically created and include:
i) name indices to reference the names of all elements and attributes,
ii) path indices to store distinct paths of the documents in the database, and
iii) document indices to reference all document nodes.
Value indices are user-defined.
They include: 
i) text indices for documents' text nodes to improve the performance of exact and range queries,
ii) attribute indices to speed up comparisons on attribute values, 
iii) token indices to improve the multi-token attribute values, and 
iv) full-text indices to normalized tokens of text nodes and speed up queries which contain text expressions.

\subsubsection*{eXist-db}

eXist-db~\cite{Meier2003} is a XDBMS implemented in Java that stores documents in the XML format.
It stores data in-memory using Document Object Model (DOM) trees.

Although eXist-db does not have support for database-level transaction control, it has transactions internally, transparent to the user, and also has a persistent journal that is used to ensures the durability and consistency of the stored data.
The database consistency is done automatically or using a sanity checker to detect the inconsistencies or damages in the core database files~\cite{Siegel2014}.

eXist-db supports data primary-secondary replication, thus allowing applications to be distributed over multiple servers through the use of Java Message Service (JMS) API.
Although replication is available, data partitioning or sharding and distributing queries across multiple servers are not.

eXist-db provides CRUD operations and ad-hoc queries for filtering and aggregation using XQuery 3.1 and XPath 3.1~\cite{Grun2009}.
Unfortunately, it does not have the MapReduce functionality, which would offer more flexibility to the aggregation queries.

eXist-db supports four types of indices~\cite{eXistdb}: 
i) range indices that provide range and field-based searches, 
ii) text indices for full-text search, 
iii) n-gram indices for improving the performance of n-gram search, and 
iv) spatial indices for querying data using geometric characteristics, although this feature is currently experimental.

\subsubsection*{Sedna}

Sedna is an XDBMS written in C that stores documents in the XML format~\cite{Fomichev2006}.
Sedna provides ACID transactions, indexing, and persistent storage~\cite{sedna}.
In uses the main memory to improve query performance~\cite{Taranov2010}.
Replication and partitioning are not implemented in Sedna.

Like the other XDBMSes, Sedna provides CRUD operations and ad-hoc queries for filtering and aggregation using XQuery 1.1 and XPath 2.0.
However, it does not provide MapReduce functionality in working with these queries.

Value indices are used to index elements' content and attributes.
Full-text indices can be created in Sedna to facilitate full-text search using XQuery.

\subsection{JDBMSes}

DODBMSes are designed for storing, retrieving, managing, and processing semi-structured data in the form of document.
With the rise of the NoSQL movement, multiple DODBMS solutions, both proprietary and open-source, have been implemented.
An important subcategory of these systems is JDBMS, which consists of systems that use the JSON format for document encoding.
For our comparison, we choose three of the more popular and open source JDBMSes\footnote{DB-Engines ranking \url{https://db-engines.com/en/ranking/document+store}}: MongoDB, CouchDB, and Couchbase.

\subsubsection*{MongoDB}

MongoDB is a DODBMS developed in C++ that focuses on combining the critical capabilities of RDBMSes with the innovations of NoSQL DBMSes.
MongoDB uses a flexible, dynamic schema to store data.
A record is stored in a document and multiple documents are stored in a collection.
Documents in a collection do not necessarily have the same structure and so the number of attributes and their data type can differ from one record to another.
In practice documents usually model objects from a high-level programming language.
Although the database allows documents with a different number of attributes and different data types for the same attributes, records have almost the same structure in a collection~\cite{Banker2016}.

MongoDB stores the data in BSON documents.
A BSON is a binary-encoded serialization of JSON-like documents.
This format is easily parsed and lightweight with respect to the overhead needed to store data.

Transactions in MongoDB respect the BASE (Basically Available, Soft state, Eventual consistency) transaction model which ensures that all the modification operations will propagate on all the nodes in an asynchronous way.
MongoDB uses Causal Consistency that enables operations to logically depend on preceding operations~\cite{MongoDB} and in-memory functionalities to improve the query execution time.
Furthermore, this JDBMS supports multi-document transactions with ACID data integrity guarantees.

To achieve redundancy and data availability, MongoDB uses Replica Sets for primary-secondary replication.
A replica set is a group of MongoDB instances that store the same dataset.
To partition the data and distribute it across multiple machines, MongoDB uses Sharding.
Sharding is a horizontal scaling mechanism that partitions and balances the data on multiple nodes or replica sets.

MongoDB supports CRUD operations and ad-hoc querying through the use of a JavaScript API available in the MongoDB client.
The Aggregation Pipeline framework is a multi-stage pipeline that transforms documents into aggregated results using the concepts of data processing pipelines.
Aggregation can also be achieved using the MapReduce framework.

MongoDB supports primary and secondary indexing.
These indices can be a single field, compound (multikey), geospatial, hashed, and text.
Text indices enable full-text search.

\subsubsection*{CouchDB}

CouchDB is an open-source DODBMS developed in Erlang that provides a schema-free model for storing self-contained data using the JSON format~\cite{CouchDB}.

Transactions in CouchDB respect document-level ACID properties with Multi-Versioning Concurrency Control (MVCC)~\cite{Anderson2010}.
CouchDB relies on Eventual Consistency together with incremental replication to maintain the data consistency.
CouchDB does not provide in-memory capabilities.
CouchDB provides primary-primary and primary-secondary asynchronous replication.
Sharding is used to distribute horizontally in a cluster the copies of each replica~\cite{Holt2011}.
To resolve inconsistencies, CouchDB uses a conflict-flagging mechanism.

CouchDB supports CRUD operations and ad-hoc querying using a JavaScript API called Mango.
For aggregation, CouchDB provides Views and MapReduce functionalities~\cite{Manyam2012}.
Indexing in CouchDB is achieved through the use of views.
CouchDB provides two types of indices: JSON and text for full-text search support.

\subsubsection*{Couchbase}

Couchbase is a highly-scalable DODBMS that stores documents using the JSON encoding.
It offers high availability, horizontal scaling, and high transaction throughput~\cite{Brown2012}.

Transactions in Couchbase respect the ACID properties and rely on Eventual Consistency and Immediate Consistency.
Couchbase has in-memory capabilities and keeps records into buckets.
The buckets are of the following type
i) Couchbase buckets used to store data persistently and in-memory,
ii) Ephemeral buckets used when persistence is not required, and
iii) Memcached buckets used to cache frequently-used data and minimize the number of queries a database-server must perform.

Couchbase uses a shared-nothing architecture and provides primary-primary and primary-secondary as well as partitioning through the use of sharding.
Couchbase scales horizontally in a cluster.

Ad-hoc data querying is achieved using a JavaScript API or a SQL-like language, i.e., N1QL (Non-1NF Query Language)~\cite{Vohra2015}.
These languages enable Couchbase to have OLTP (Online Transaction Processing) CRUD operations and ETL (Extract Transform Load) capabilities~\cite{Hubail2019}.

JavaScript MapReduce Views can be developed and stored on the server-side to specify complex indexing and aggregation queries~\cite{Couchbase}.

Couchbase provides multiple types of indices:~\cite{Couchbase}
i) composite indices to index multiple attributes,
ii) covering indices to index the information needed for querying without accessing the data,
iii) filtered (partial) indices to index a subset of the data used by the WHERE clause,
iv) function-based indices that compute the value of an expression over a range of documents,
v) sub-document indices to index embedded structures,
vi) incremental MapReduce views to index the results of complex queries that perform sorting and aggregation to support real-time analytics over very large datasets,
vii) spatial views using Spatial MapReduce to index multi-dimensional numeric data, and
viii) full-text indices used for full-text search capabilities.

\subsection{DODBMSes Comparison}

Table~\ref{tbl:comp} summarizes the main features of the presented databases.
BaseX, Sedna and Couchbase offer ACID compliant transactions in comparison with MongoDB that offers BASE compliant multi-document isolation transactions and CouchDB that offers document-level ACID with MVCC transactions.
XDBMSes support transaction consistency while MongoDB and CouchDB support casual consistency and eventual consistency, respectively.
Couchbase supports both eventual and immediate consistency.
A disadvantage of XDBMSes is that they do not have replication or partitioning mechanisms, except for eXist-db which offers primary-secondary replication.
An advantage of XDBMSes is the use of XQuery and XPath for querying the data which makes ad-hoc querying an easy task.
Although XDBMSes support aggregation queries, they do not provide MapReduce frameworks as a result of the lack of distribution capabilities.
Another advantage of XDBMSes is that they offer different types of indices, including text indices for full-text search.
As can be seen from Table~\ref{tbl:comp}, the chosen JDBMS solutions also offer different types of indices, but in addition to JDBMS, the one used in XDBMS systems can also be added on properties and paths, not only on keys and values.

\begin{table*}[!ht]
\centering
\caption{DODBMS comparison}
\label{tbl:comp}
\resizebox{\textwidth}{!}{
\begin{tabular}{l|l|l|l|l|l|l|}
\cline{2-7}
& \multicolumn{1}{c|}{\textbf{BaseX}} & \multicolumn{1}{c|}{\textbf{eXist-db}} & \multicolumn{1}{c|}{\textbf{Sedna}} & \multicolumn{1}{c|}{\textbf{MongoDB}} & \multicolumn{1}{c|}{\textbf{CouchDB}} & \multicolumn{1}{c|}{\textbf{Couchbase}} \\ \hline
\multicolumn{1}{|l|}{\textbf{DBMS type}}      & XDBMS & XDBMS & XDBMS & JDBMS & JDBMS & JDBMS \\ \hline
\multicolumn{1}{|l|}{\textbf{Data format}}    & XML   & XML   & XML   & BSON (Binary JSON) & JSON & JSON\\ \hline
\multicolumn{1}{|l|}{\textbf{Implementation}} & Java & Java & C & C++ & Erlang & C/C++, Go, Erlang \\ \hline
\multicolumn{1}{|l|}{\textbf{Transaction}}    & ACID & Isolation safe & ACID  & \begin{tabular}[c]{@{}l@{}}BASE\\Multi-document isolation\end{tabular} & \begin{tabular}[c]{@{}l@{}}Document-level ACID\\with MVCC\end{tabular} & ACID   \\ \hline
\multicolumn{1}{|l|}{\textbf{Consistency}}    & Transaction Consistency  & \begin{tabular}[c]{@{}l@{}}Automatic consistency\\Sanity checker\end{tabular}                   & Transaction Consistency & Causal Consistency & Eventual Consistency  & \begin{tabular}[c]{@{}l@{}}Eventual Consistency\\Immediate Consistency\end{tabular}
\\ \hline
\multicolumn{1}{|l|}{\textbf{In-memory}} & Yes & Yes & Yes & Yes & No & Yes \\ \hline
\multicolumn{1}{|l|}{\textbf{Replication}} & No & Primary-Secondary & No & Primary-Secondary & \begin{tabular}[c]{@{}l@{}}Primary-Primary\\Primary-Secondary\end{tabular} & \begin{tabular}[c]{@{}l@{}}Primary-Primary\\Primary-Secondary\end{tabular} \\ \hline
\multicolumn{1}{|l|}{\textbf{Partitioning}} & No & No & No & Sharding & Sharding & Sharding \\ \hline
\multicolumn{1}{|l|}{\textbf{Ad-hoc queries}} & \begin{tabular}[c]{@{}l@{}}XQuery 3.1\\XPath 3.1\end{tabular} & \begin{tabular}[c]{@{}l@{}}XQuery 3.1\\XPath 3.1\end{tabular} & \begin{tabular}[c]{@{}l@{}}XQuery 1.0\\XPath 2.0\end{tabular} & JavaScript & Mango & \begin{tabular}[c]{@{}l@{}}N1QL\\JavaScript\end{tabular} \\ \hline
\multicolumn{1}{|l|}{\textbf{MapReduce}} & No & No & No & Yes & Yes & Yes \\ \hline
\multicolumn{1}{|l|}{\textbf{Secondary indices}} & Yes & Yes & Yes & Yes & Yes & Yes \\ \hline
\multicolumn{1}{|l|}{\textbf{Geospatial indices}} & No & No & Yes & Yes & Yes & Yes \\ \hline
\multicolumn{1}{|l|}{\textbf{Text indices}} & Yes & Yes & Yes & Yes & Yes & Yes \\ \hline
\end{tabular}
}
\end{table*}

\section{Benchmark specifications}\label{sec:specs}

\subsection{Data Model}

\begin{figure*}[!ht]
    \centering
    \includegraphics[width=0.9\columnwidth]{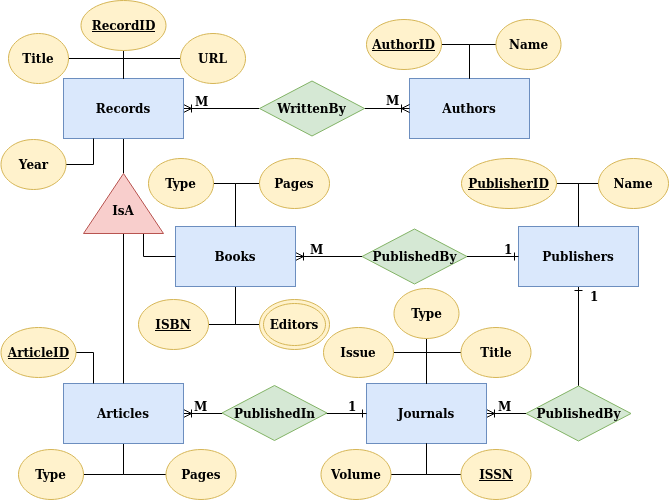}
    \caption{Database entity-relational diagram}
    \label{fig:dblper}
\end{figure*}

For our benchmark, we proposed a heterogeneous entity-relationship schema that can be easily expanded with more complex relationships and new entities.
Figure~\ref{fig:dblper} presents the proposed schema.
The model's entities are described below.

\begin{itemize}
\item 
    \textit{Authors} is the entity that stores information about authors.
    Besides the unique identifier for each author \textit{AuthorID}, the attribute \textit{Name} is used for storing the name of each author.
\item 
    \textit{Records} contains information about the published work of one or more authors.
    It stores the \textit{Title}, the \textit{URL} for quick access on the web, and the publishing \textit{Year}.
    The many-to-may relationship \textit{WrittenBy} correlates each record with the authors.
    A record can be either published as a book (or book chapter) or as an article (conference or journal).
    The relationship \textit{IsA} is used for denoting the sub-type of a record.
\item
    \textit{Books} is the first sub-type of a record.
    This entity stores the following information: 
    i) the unique book identifier \textit{ISBN}, 
    ii) the pages of a record using the attribute \textit{Pages}, 
    iii) the book editors using the multi-variate attribute \textit{Editors}, and 
    iv) the type of a record of this sub-type, i.e., book or book chapter, using the attribute \textit{Type}.
    The one-to-many relationship \textit{PublishedBy} is used to correlate each record of sub-type \textit{Book} to a \textit{Publisher}.
\item 
    \textit{Articles} is the second sub-type of a record.
    Besides the unique identifier of a record in this sub-type, the entity \textit{Articles} stored information about 
    i) the pages of a record using the attribute \textit{Pages}, and 
    ii) the type of a record of this sub-type, i.e., conference or journal article, using the attribute \textit{Type}.
    The one-to-many relationship \textit{PublishedIn} is used to correlate each article to a journal.
\item 
    \textit{Journals} entity stores information about an article publication venue.
    The attributes are: 
    i) \textit{ISSN} used as the unique identifier, 
    ii) \textit{Type} used to determine if the publication is a journal, proceedings, or special issue, 
    iii) \textit{Title} used for keeping the title of the journal or the conference name,
    iv) \textit{Volume} used to store the number of years since the first publication, and
    v) \textit{Issue} used to store how many times the journal has been published during a year.
    The one-to-many relationship \textit{PublishedBy} is used to correlate each record of sub-type \textit{Journal} to a \textit{Publisher}.
\item 
    \textit{Publishers} is the entity that stores a unique identifier and the \textit{Name} of a publishing house.
\end{itemize}

\subsection{Workload Model}\label{subsec:workload}

The workload model follows two analysis directions:
i) selection queries for filtering the corpus and extract subsamples, and 
ii) aggregation queries for creating reports.

For the selection queries, a constraint $c^{i}_{1} = contains(Records.Title, t_{i})$ is used to extract the most relevant records that are contained in the title of a given set of terms.
The constraint $c^{i}_{1}$ utilizes the $contains(\cdot,\cdot)$ function, which verifies if a substring $t_{i} \in \{ t | t \in vocabulary \}$ belongs to a string.
In this case, the $vocabulary$ is the set of terms extracted from each title using Tokenization.

Aggregation queries are used to create reports about the publishing activity of each author.
These reports are created by counting the number of published records using attributes for grouping.
To achieve this, we apply the aggregation operator $\gamma_L$ with $L=(F, G)$, where $F$ is the list of aggregation functions, and $G$ is the list of attributes in the GROUP BY clause.
We use the \textit{Authors.Name} attribute in the GROUP BY clause to create an overview report of the publication activity for each author over his/her entire academic life.
To determine the publishing patterns by year of each author, we use the \textit{Records.Year} attribute that adds a time dimension to the previous report.
For a more in-depth analysis of each published topic by author, we also use the $c^{i}_{1}$ constraint to filter the dataset by keywords before counting the number of articles.

\section{Benchmark Implementation}\label{sec:impl}

\subsection{Database Design}

\begin{figure*}[!ht]
    \centering
    \begin{subfigure}{0.45\columnwidth}
        \centering
        \includegraphics[width=\columnwidth,height=0.7\columnwidth]{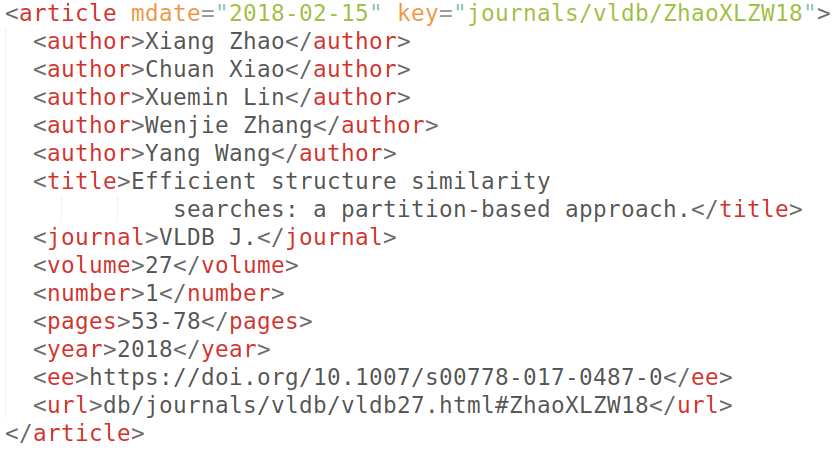}
        \caption{XML Document}
        \label{fig:xml}
    \end{subfigure}
    \begin{subfigure}{0.45\columnwidth}
        \centering
        \includegraphics[width=\columnwidth,height=0.7\columnwidth]{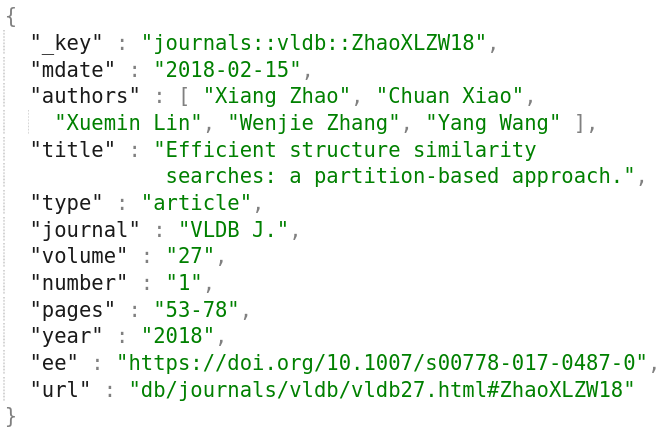}
        \caption{JSON Document}
        \label{fig:json}
    \end{subfigure}
    \caption{Document representation in XML and JSON}
    \label{fig:docs}
\end{figure*}

The conceptual entity-relational diagram described in Section~\ref{sec:specs} must be translated into the XML and JSON formats (Figure~\ref{fig:docs}).
For the XML representation (Figure~\ref{fig:xml}), the attributes of entities are directly encoded in the elements' names, e.g., the \textit{Article.Type} is directly encoded into the \textit{journal} label.
In the case of the \textit{Authors} entity, the records associated with the article are presented as multiple tags with the same name, i.e., \textit{author}.
For the JSON representation, the \textit{Authors} entity becomes a list of values, i.e., the label \textit{authors}.
The information regarding an article is stored directly in the document using labels, e.g., type, publication year, etc.
Using this representation, both schemes are greatly simplified and the need of relationships between entities disappears.

\subsection{Query Description}

The proposed benchmark features nine queries with different complexity and selectivity, i.e., $Q_1$ to $Q_9$.
The first five queries are used to filter the dataset based on different constraints.
Whereas, the last four queries are used to filter and group the data in order to obtain aggregated results.

\subsubsection{Selection Queries}

The first set of queries selects the records that respect a given constraint.

The first query ($Q^{i}_{1}$) uses the constraint $c^{i}_{1}$ to extract the documents which contain in their title a certain given term $t_i$ (Equation~\eqref{eq:q1}).
The projection for the query, which specifies the set of selected attributes following the query execution, is $\Pi_{1} = \{ Records.Title \}$.

\begin{equation}\label{eq:q1}
    Q^{i}_1 = \pi_{\Pi_{1}} (\sigma_{c^{i}_{1}}(Records))
\end{equation}

The second query ($Q^{ij}_{2}$) extracts the records that contain in their title two terms (Equation~\eqref{eq:q2}).
It uses the constraint $c^{s}_{1}$, $s \in \{i, j \}$ with $i \neq j$.
The query is written using the INTERSECTION operator between the results returned by $Q^{i}_{1}$ for term $t_{i}$ and $Q^{j}_{1}$ for term $t_{j}$.
Due to the nature of the filtering condition, we can concatenate the separate conditions to create a single conditional expression using the \textit{and} logical operator ($\wedge$), i.e., $c^{i}_{1} \wedge c^{j}_{1}$.
As in the case of the first query, the projection remains $\Pi_{1}$.

\begin{equation}\label{eq:q2}
\begin{split}
    Q^{ij}_{2} 
               & = Q^{i}_{1} \cap Q^{j}_{1} \\
               & = \pi_{\Pi_{1}} (\sigma_{c^{i}_{1}}(Records)) \cap \pi_{\Pi_{1}}(\sigma_{c^{j}_{1}}(Records)) \\
               & = \pi_{\Pi_{1}} (\sigma_{c^{i}_{1} \wedge c^{j}_{1}}(Records)) \\ 
\end{split}
\end{equation}

$Q^{ij}_{3}$ extracts the records that contain in their title at least one of the terms given through the $c^{i}_{1}$ or $c^{j}_{1}$ constraints, with $i \neq j$ (Equation~\eqref{eq:q3}).
The query is written using the UNION operator between the results returned by $Q^{i}_{1}$ for term $t_{i}$ and $Q^{j}_{1}$ for term $t_{j}$.
The projection remains $\Pi_{1}$.
As for query $Q^{ij}_{2}$, the conditions can be concatenated to create a single conditional expression using the \textit{or} logical operator ($\vee$), i.e., $c^{i}_{1} \vee c^{j}_{1}$.

\begin{equation}\label{eq:q3}
\begin{split}
    Q^{ij}_{3} & = Q^{i}_{1} \cup Q^{j}_{1} \\
               & = \pi_{\Pi_{1}} (\sigma_{c^{i}_{1}}(Records)) \cup \pi_{\Pi_{1}} (\sigma_{c^{j}_{1}}(Records)) \\
               & = \pi_{\Pi_{1}} (\sigma_{c^{i}_{1} \vee c^{j}_{1}}(Records)) \\
\end{split}
\end{equation}

The fourth query ($Q_4$) filters the \textit{Records} entity and extracts the documents that contain in their title the terms $t_{i}$, $t_{j}$, and $t_{k}$ (Equation~\ref{eq:q4}).
As for the previous queries, the projection attributes are given using $\Pi_{1}$.
The query is written using the INTERSECTION operator between the results obtained by $Q^{i}_{1}$, $Q^{j}_{1}$, and $Q^{k}_{1}$ for terms $t_{i}$, $t_{j}$, and $t_{k}$ respectively.
Due to the nature of the filtering conditions, they can be concatenated into one constraint $c^{i}_{1} \wedge c^{j}_{1} \wedge c^{k}_{1}$.

\begin{equation}\label{eq:q4}
\begin{split}
    Q^{ijk}_{4} & = Q^{i}_{1} \cap Q^{j}_{1} \cap Q^{k}_{1} \\
                & = \pi_{\Pi_{1}} (\sigma_{c^{i}_{1}}(Records)) \cap \pi_{\Pi_{1}}(\sigma_{c^{j}_1}(Records)) \cap \pi_{\Pi_{1}}(\sigma_{c^{k}_{1}}(Records)) \\
                & = \pi_{\Pi_{1}} (\sigma_{c^{i}_{1}}(Records) \cap \sigma_{c^{j}_{1}}(Records) \cap \sigma_{c^{k}_{1}}(Records)) \\
                & = \pi_{\Pi_{1}} (\sigma_{c^{i}_{1} \wedge c^{j}_{1} \wedge c^{k}_{1}}(Records)) \\
\end{split}    
\end{equation}

The last selection query ($Q_{5}$) extracts the documents that contain in their title one or more of the searched terms $t_{s}$, $s \in \{i, j, k \}$ with $i \neq j \wedge i \neq k \wedge j \neq k$.
The query is written using the UNION operator between the results obtained by each $Q^{s}_{1}$ for $t_{s}$ terms.
The nature of the filtering constraints permit the query to be written using one constraint $c^{i}_{1} \vee c^{j}_{1} \vee c^{k}_{1}$ and the projection $\Pi_{1}$ (Equation~\ref{eq:q5}).

\begin{equation}\label{eq:q5}
\begin{split}
    Q^{ijk}_{5} & = Q^{i}_{1} \cup Q^{j}_{1} \cup Q^{k}_{1} \\
                & = \pi_{\Pi_{1}} (\sigma_{c^{i}_{1}}(Records)) \cup \pi_{\Pi_{1}} (\sigma_{c^{j}_{1}}(Records)) \cup \pi_{\Pi_{1}} (\sigma_{c^{k}_{1}}(Records)) \\
                & = \pi_{\Pi_{1}} (\sigma_{c^{i}_{1}}(Records) \cup \sigma_{c^{j}_{1}}(Records) \cup \sigma_{c^{k}_{1}}(Records)) \\
                & = \pi_{\Pi_{1}} (\sigma_{c^{i}_{1} \vee c^{j}_{1} \vee c^{k}_{1}}(Records)) \\
\end{split}
\end{equation}

\subsubsection{Aggregation Queries}

The last four queries use aggregation to count the number of articles using different filtering constraints and attributes in the GROUP BY clause.

The sixth query ($Q_{6}$) uses aggregation to determine the number of articles written by each author (Equation~\eqref{eq:q6}).
It uses a JOIN operation between the \textit{Records} and \textit{Authors} entities.
Because there is a many-to-many relationship between the two entities, the JOIN also traverses \textit{WrittenBy}.
The projection attributes are $\Pi_{6} = \{Author.Name, count\}$.
To determine the number of articles for each author, we use the aggregation operator $\gamma_{L_{6}}$, where $L_{6}=(F_{6}, G_{6})$.
The list of aggregation functions is given by $F_{6}$, while the set of attributes in the GROUP BY clause is given by $G_{6}$.
The list of aggregation functions is $F_{6} = \{ count(Records.RecordID) \}$, where the $count$ is the counting aggregation function.
The set of attributes in the GROUP BY clause is $G_{6} = \{ Authors.Name \}$.

\begin{equation}\label{eq:q6}
    Q_{6} = \pi_{\Pi_{6}}(\gamma_{L_{6}}(Authors \bowtie Records))
\end{equation}

The seventh query ($Q_{7}$) counts the number of articles published by an author for each year (Equation~\eqref{eq:q7}).
The query makes use of a JOIN operation between the \textit{Records} and \textit{Authors} entities, as in the case of query $Q_{6}$.
The projection uses the following attributes $\Pi_{7} = \{ Author.Name, Record.Year, count \}$.
To determine the number of articles written in a year by each author, we use the aggregation operator $\gamma_{L_{7}}$, where $L_{7}=(F_{7}, G_{7})$.
For query $Q_{7}$, the list of aggregation functions is given by $F_{7}$, while the set of attributes in the GROUP BY clause is given by $G_{7}$.
The list of aggregation functions is $F_{7} = \{ count(Records.RecordID) \}$, where the $count$ is the counting function used for determining the number of articles written in a year by each author.
The set of attributes in the GROUP BY clause is $G_{7} = \{ Authors.Name, Records.Year \}$.

\begin{equation}\label{eq:q7}
    Q_{7} = \pi_{\Pi_{7}}(\gamma_{L_{7}}(Authors \bowtie Records))
\end{equation}

The eighth query ($Q_{8}$) extracts the documents that contain in their title all of the searched terms, and then it counts the number of articles grouped by author and year.
As in the case of $Q_{6}$, the JOIN operation is between the \textit{Records} and \textit{Authors} entities.
The query is written using the INTERSECTION operator.
The filtering is done using the constraints $c^{i}_{1}$, $c^{j}_{1}$, $c^{k}_{1}$ which ensures that the title contains all terms $t_{i}$, $t_j$, and $t_k$ with $i \neq j \wedge i \neq k \wedge j \neq k$.
The projection attributes and the aggregation operator remains the same as in the case of $Q_{7}$, i.e., $\Pi_{7}$ and $\gamma_{L_{7}}$.
Due to the nature of the filtering conditions, the query can be rewritten using only one constraint $c^{i}_{1} \wedge c^{j}_{1} \wedge c^{k}_{1}$.

\begin{equation}\label{eq:q8}
\begin{split}
    Q_{8} 
          & = \pi_{\Pi_{7}}(\gamma_{L_{7}}(\sigma_{c^{i}_{1}}(Records \bowtie Authors) \cap \sigma_{c^{j}_{1}}(Records \bowtie Authors) \cap \sigma_{c^{k}_{1}}(Records \bowtie Authors))) \\
          & = \pi_{\Pi_{7}}(\gamma_{L_{7}}(\sigma_{c^{i}_{1} \wedge c^{j}_{1} \wedge c^{k}_{1}}(Records \bowtie Authors))) \\
\end{split}    
\end{equation}

The last query ($Q_9$) extracts the documents that contain in their title one or more of the searched terms $t_s$, $s \in \{i, j, k\}$ and $i \neq j \wedge i \neq k \wedge j \neq k$, by filtering through the use of constraint $c^{s}_{1}$.
The JOIN operator is used once again between the \textit{Records} and \textit{Authors} entities, as in the case of $Q_{6}$.
The projection attributes and the aggregation operator remain the same as in the case of $Q_{7}$, i.e., $\Pi_{7}$ and $\gamma_{L_{7}}$.
The filtering constraints $c^{i}_{1}$, $c^{j}_{1}$, $c^{k}_{1}$ are applied on the \textit{Records} entity.
The query uses the UNION operator between the relationship obtained after filtering.
Due to the nature of the filtering, the query can be rewritten using one constraint $c^{i}_{1} \vee c^{j}_{1} \vee c^{k}_{1}$.

\begin{equation}\label{eq:q9}
\begin{split}
    Q_{9} 
        & = \pi_{\Pi_{7}}(\gamma_{L_{7}}(\sigma_{c^{i}_{1}}(Records \bowtie Authors) \cup \sigma_{c^{j}_{1}}(Records \bowtie Authors) \cup \sigma_{c^{k}_1}(Records \bowtie Authors))) \\
        & = \pi_{\Pi_{7}}(\gamma_{L_{7}}(\sigma_{c^{i}_{1} \vee c^{j}_{1} \vee c^{k}_{1}}(Records \bowtie Authors))) \\
\end{split}
\end{equation}

\section{Experiments}\label{sec:experiments}

\subsection{Experimental Conditions}\label{subsec:conditions}

All tests were run on an IBM System x3550 M4 with 64GB of RAM, and an Intel(R) Xeon(R) CPU E5-2670 v2 @ 2.50GHz.
The XDBMSes used for benchmarking are BaseX, eXist-db, and Sedna.
For comparison reasons we also use three JDBMSes: MongoDB, CouchDB, and Couchbase.
We chose these DODBMSes because they are free to uses and because their licenses do not forbid benchmarking.

The versions of the deployed DODBMSes are listed in Table~\ref{tbl:dbs}.
The proposed benchmark, the results, and the used dataset are publicly available on-line\footnote{GitHub Sources~\url{https://github.com/cipriantruica/The-Forgotten-DODBMSes}}.

\begin{table}[!ht]
\centering
\caption{Benchmarked DODBMSes}
\label{tbl:dbs}
\begin{tabular}{|l|l|}
\hline
\textbf{DODBMS} & \textbf{Version} \\ \hline
BaseX             & 9.3.3          \\ \hline
eXist-db          & 5.2.0          \\ \hline
Sedna             & 3.5            \\ \hline
MongoDB           & 4.2.7          \\ \hline
CouchDB           & 3.1.0          \\ \hline
Couchbase         & 6.5.1          \\ \hline
\end{tabular}
\end{table}

As the chosen XDBMS solutions do not have partitioning, we could not distribute them.
Therefore, we deployed and tested them on a single instance environment.
Moreover, for comparison reasons, we also used a single instance environment for MongoDB, CouchDB, and Couchbase.

The query parameterization is presented in Table~\ref{tbl:params}.
Each term $t_i$ ($i = \overline{1, 3}$) is used for filtering the records through the constraint $c^{(i)}_{1}$.
Thus for the first set of queries, i.e., $Q^{i}_{1}$, $Q^{ij}_{2}$, and $Q^{ijk}_{3}$, the $i$, $j$, and $k$ indices ($i \neq j \wedge i \neq k \wedge j \neq k$) represent the $i'\in \overline{1,3}$ index of the $t_{i'}$ used for filtering.

\begin{table}[!ht]
\centering
\caption{Query parameter values}
\label{tbl:params}
\begin{tabular}{|l|c|}
\hline
Parameter & Value    \\ \hline
$t_{1}$   & database \\ \hline
$t_{2}$   & text     \\ \hline
$t_{3}$   & mining   \\ \hline
\end{tabular}
\end{table}

\subsection{Dataset}

The experiments are performed on 6\,150\,738 records extracted from DBLP\footnote{DBLP \url{http://dblp.org/}}.
The initial dataset is split into $4$ different subsets to test the scalability of each DODBMS w.r.t. the number of records.
These subsets contain 768\,842, 1\,537\,685, 3\,075\,369, and 6\,150\,738 records, respectively.
Each subset allows scaling experiments and are associated with a scale factor $SF$ parameter, where $SF = \{0.125, 0.25, 0.5, 1 \}$.
Table~\ref{tab:dataset} presents the size of the $4$ subsets, both as raw data and the resulting DODBMS collection dimension.

\begin{table*}[!ht]
\centering
\caption{Dataset}
\label{tab:dataset}
\resizebox{\textwidth}{!}{
\begin{tabular}{|c|r|r|r|r|r|r|r|r|r|}
\hline
\multicolumn{1}{|c|}{\textbf{SF}} & \multicolumn{1}{c|}{\textbf{\begin{tabular}[c]{@{}c@{}}No. \\ Records\end{tabular}}} & \multicolumn{1}{c|}{\textbf{\begin{tabular}[c]{@{}c@{}}Raw\\ XML\end{tabular}}} & \multicolumn{1}{c|}{\textbf{\begin{tabular}[c]{@{}c@{}}Raw\\ JSON\end{tabular}}} & \multicolumn{1}{c|}{\textbf{\begin{tabular}[c]{@{}c@{}}BaseX\\ DB size\end{tabular}}} & \multicolumn{1}{c|}{\textbf{\begin{tabular}[c]{@{}c@{}}eXist-db\\ DB size\end{tabular}}} & \multicolumn{1}{c|}{\textbf{\begin{tabular}[c]{@{}c@{}}Sedna \\ DB size\end{tabular}}} & \multicolumn{1}{c|}{\textbf{\begin{tabular}[c]{@{}c@{}}MongoDB\\ DB size\end{tabular}}} & \multicolumn{1}{c|}{\textbf{\begin{tabular}[c]{@{}c@{}}CouchDB\\ DB size\end{tabular}}} & \multicolumn{1}{c|}{\textbf{\begin{tabular}[c]{@{}c@{}}Couchbase\\ DB size\end{tabular}}} \\ \hline
0.125 &    768\,842 & 0.38GB & 0.34GB & 0.53GB & 0.44GB &  1.78GB & 0.17GB & 0.43GB & 0.44GB \\ \hline
0.25  & 1\,537\,685 & 0.75GB & 0.67GB & 1.05GB & 0.86GB &  3.36GB & 0.33GB & 0.85GB & 0.92GB \\ \hline
0.5   & 3\,075\,369 & 1.51GB & 1.36GB & 2.09GB & 1.74GB &  6.71GB & 0.67GB & 1.69GB & 1.78GB \\ \hline
1     & 6\,150\,738 & 2.25GB & 2.06GB & 3.14GB & 2.59GB & 10.17GB & 1.02GB & 2.81GB & 3.02GB \\ \hline
\end{tabular}
}
\end{table*}

For all the XDBMSes as well as for CouchDB and Couchbase, we can observe that database size is larger than the raw dataset.
This increase is a direct result of the overhead required by the DODBMSes to manage and store the data.
MongoDB uses compression mechanisms, which in turn manage to decrease the database size by minimizing the overhead.

\subsection{Query Implementation}

Data are stored within each DODBMS using a denormalized schema; thus, one-to-many and many-to-many relationships are encapsulated inside the same document.
To achieve denormalization, JDBMSes employ nested documents, lists, and lists of nested documents, while XDBMSes use the hierarchical structure of the XML format.
To normalize the information and apply filtering and aggregation operations and functions, we use the native syntax, operators, query language clauses, and frameworks provided by each DODBMS.
Table~\ref{tbl:queries} presents the implementation language and operators.

\begin{table*}[!ht]
\centering
\caption{Filtering and aggregation queries}
\label{tbl:queries}
\begin{tabular}{|l|l|l|}
\hline
\textbf{Database} & \textbf{Filtering Query} & \textbf{Aggregation Queries}                \\ \hline
BaseX             & XQuery 3.1                        & XQuery 3.1 syntax for sorting and grouping           \\ \hline
eXist-db          & XQuery 3.1                        & XQuery 3.1 syntax for sorting and grouping           \\ \hline
sedna             & XQuery 1.0                        & XQuery 1.1 syntax for sorting and grouping           \\ \hline
MongoDB           & JavaScript                        & JavaScript Aggregation Pipeline with \texttt{unwind} operator \\ \hline
CouchDB           & JavaScript/Mango                        & JavaScript/Mango Materialized Views                        \\ \hline
Couchbase         & N1QL                              & N1QL with \texttt{UNNEST} clause                            \\ \hline
\end{tabular}
\end{table*}

For the XDBMSes, we implemented the queries using XQuery.
The aggregation queries for BaseX and eXist-db use the XQuery 3.1 syntax for sorting and grouping, i.e., \texttt{FOR ... WHERE ... GROUP BY ... ORDER BY ...}.
For Sedna, we use the XQuery 1.1 syntax for sorting and grouping, i.e., \texttt{FOR ... WHERE ... LET ... ORDER BY ...}.
We used the native Command Line Interfaces to run these queries.

The aggregation queries in MongoDB are implemented using its Aggregation Pipeline framework.
To deal with nested documents, the \texttt{unwind} operator is used to flatten an array field of nested documents.
This operator is useful when trying normalize the one-to-many and many-to-many which trough denormalization are stored in the JSON format as nested documents or lists of nested documents.
We used the native Command Line Interfaces to run these queries.

CouchDB uses Materialized Views for aggregation and to deal with nested and list of nested documents.
These views are implemented using CouchDB's MapReduce framework.
The mapper function is used to flatten nested documents and filter the field.
The reducer function is used for applying an aggregation function and returning the final result.
We used cURL to run these queries.

To manipulate nested array in Couchbase, N1QL offers developers the UNNEST clause.
This clause is used to flatten the arrays in the parent document.
Thus, the UNNEST clause conceptually performs a JOIN operation between nested arrays and the parent document.
As data are stored using the JSON format, the JOIN operation increases the runtime and decreases the overall retrieval performance.
For Couchbase, we used the native Command Line Interfaces to run these queries.

\subsection{Query Selectivity}

Selectivity, i.e., the amount of retrieved data ($n(Q)$) w.r.t. the total amount of available data ($N$), depends on the number of attributes in the WHERE and GROUP BY clauses.
The selectivity formula used for a query $Q$ is $S(Q) = 1-\frac{n(Q)}{N}$.
For the selection queries, we set $N$ equal to the cardinality of the \textit{Records} entity, i.e., $N = ||Records||$.
Table~\ref{tbl:filter_queries} presents the filtering queries' selectivity w.r.t. the $SF$.
The queries with more restrictive conditions return a smaller number of records and the selectivity is higher, e.g., $Q^{ij}_2$.
The queries with more inclusive restrictions return a higher number of records and the selectivity is lower, e.g., $Q^{ij}_3$.

\begin{table*}[!ht]
\centering
\caption{Filter queries selectivity}
\label{tbl:filter_queries}
\begin{tabular}{|c|c|c|c|c|c|c|c|c|c|c|c|}
\hline
$SF$  & $Q^{1}_{1}$ & $Q^{2}_{1}$ & $Q^{3}_{1}$ & $Q^{12}_{2}$ & $Q^{13}_{2}$ & $Q^{23}_{2}$ & $Q^{12}_{3}$ & $Q^{13}_{3}$ & $Q^{23}_{3}$ & $Q_{4}$ & $Q_{5}$ \\ \hline
0.125 & 0.992       & 0.987       & 0.993       & 0.999        & 0.999        & 0.999        & 0.980        & 0.986        & 0.980        & 0.999   & 0.974   \\ \hline
0.25  & 0.991       & 0.986       & 0.992       & 0.999        & 0.999        & 0.999        & 0.978        & 0.984        & 0.979        & 0.999   & 0.971   \\ \hline
0.5   & 0.990       & 0.982       & 0.991       & 0.999        & 0.999        & 0.999        & 0.973        & 0.982        & 0.975        & 0.999   & 0.966   \\ \hline
1     & 0.993       & 0.987       & 0.994       & 0.999        & 0.999        & 0.999        & 0.981        & 0.988        & 0.982        & 0.999   & 0.976   \\ \hline
\end{tabular}
\end{table*}

For the aggregation queries, we set $N$ equal to the number of queries returned by joining the entities \textit{Records} with \textit{Authors}, i.e., $N = || Authors \bowtie Records || $.
Table~\ref{tbl:agg_queries} shows the aggregation queries' selectivity w.r.t. the $SF$ factor.
Query $Q_{8}$ is the most restrictive query.
Because of the filtering and grouping conditions, $Q_{8}$ returns a small number of records, and its selectivity is almost equal to $1$.
The most inclusive query is $Q_{7}$, and it has a low selectivity w.r.t. $SF$.
Because of the less restrictive filtering and grouping conditions, the selectivity of this query is less than $0.45$.
The selectivity of $Q_{6}$ increases w.r.t. $SF$, meaning that the number of records returned by the query increases more gradually than the size of the corpus.

\begin{table}[!ht]
\centering
\caption{Aggregation queries selectivity}
\label{tbl:agg_queries}
\begin{tabular}{|c|c|c|c|c|}
\hline
$SF$   & $Q_{6}$ & $Q_{7}$ & $Q_{8}$ & $Q_{9}$ \\ \hline
0.125  & 0.651   & 0.256   & 0.999   & 0.974   \\ \hline
0.25   & 0.728   & 0.345   & 0.999   & 0.970   \\ \hline
0.5    & 0.797   & 0.448   & 0.999   & 0.969   \\ \hline
1      & 0.848   & 0.424   & 0.999   & 0.974   \\ \hline
\end{tabular}
\end{table}

\subsection{Performance Metrics and Execution Protocol}\label{subsec:protocol}

We use the query response time as the only metric for the benchmark.
It is symbolized for each query by $t(Q^{*}_{i} ) \forall i \in [1, 9]$.
All queries are executed $10$ times, which is sufficient according to the central limit theorem.
Additionally, all executions are warm runs, i.e., either caching mechanisms must be deactivated, or a cold run where each query must be executed once (but not taken into account in the benchmark's results) to fill in the cache.
Queries must be written in the native scripting language of the target DODBMS and executed directly inside the specified system using the command line interpreter.
Lastly, the average response time and standard deviation are computed for each $t(Q^{*}_{i})$.

\subsection{Results}

Figure~\ref{fig:q1} presents the results of $Q^{i}_{1}$ where $i =\overline{1,3}$ is used to denote the keyword $t_{i}$.
MongoDB and BaseX offer the fastest time performance among the DODBMSes that encode documents using JSON and XML, respectively, regardless of the keyword w.r.t. $SF$.
For $Q^{2}_{1}$ query which has the lowest selectivity of the three $Q^{i}_{1}$ queries, the time performance of CouchDB is with a factor of $\sim2$x faster than eXist-db w.r.t. $SF$.
The time performance of CouchDB and eXist-db for $Q^{1}_{1}$ and $Q^{3}_{1}$ tend to become the same w.r.t. $SF$, i.e., the performance difference factor between CouchDB and eXist-db at $SF=0.125$ is $\sim0.8$x which increases to $\sim0.9$x for $SF=1$.
CouchDB time performance is with a factor of $\sim1.1$x faster than Couchbase for all the $Q^{i}_{1}$ queries regardless of $SF$.
Couchbase and eXist-db have similar performance for query $Q^{3}_{1}$ and $SF=1$.
Sedna performance is almost constant regardless of query selectivity w.r.t. $SF$.
The overall best performance is achieved by MongoDB.

\pgfplotsset{height=0.70\textwidth, width=1.00\columnwidth,
/pgfplots/ybar legend/.style={
        /pgfplots/legend image code/.code={
        \draw[##1,/tikz/.cd,bar width=3pt,yshift=-0.2em,bar shift=0pt]
                plot coordinates {(0cm,0.8em)};},
},
}
\begin{figure*}[!ht]
    \centering
    \begin{subfigure}{0.5\columnwidth}
    \begin{tikzpicture}[]
        \begin{axis}[
        xmin=0,
        xmax=5,
        ymin=0,
        ymax=250,
        bar width=2pt,
        ybar = 2pt,
        xtick \empty, 
        extra x ticks={0,1,2,3,4,5}, 
        extra x tick labels={$SF$,0.125,0.25,0.5,1},
        ylabel={Response time (s)},
		legend style={anchor=north, legend columns=3, legend cell align=left, at={(0.5,1.35)}},
        ]

		\addplot+[color=blue] [error bars/.cd, y dir = both, y explicit] table [x=NO_DOCS, y=BASEX_AVG, y error = BASEX_STD, col sep=comma] {q1_kw1.csv};
        \addlegendentry{BaseX};
        \addplot+[color=red] [error bars/.cd, y dir = both, y explicit] table [x=NO_DOCS, y=EXISTDB_AVG, y error = EXISTDB_STD, col sep=comma] {q1_kw1.csv};
        \addlegendentry{eXist-db};
        \addplot+[color=green] [error bars/.cd, y dir = both, y explicit] table [x=NO_DOCS, y=SEDNA_AVG, y error = SEDNA_STD, col sep=comma] {q1_kw1.csv};
        \addlegendentry{Sedna};
        \addplot+[color=black] [error bars/.cd, y dir = both, y explicit] table [x=NO_DOCS, y=MONGODB_AVG, y error = MONGODB_STD, col sep=comma] {q1_kw1.csv};
        \addlegendentry{MongoDB};
        \addplot+[color=pink] [error bars/.cd, y dir = both, y explicit] table [x=NO_DOCS, y=COUCHDB_AVG, y error = COUCHDB_STD, col sep=comma] {q1_kw1.csv};
        \addlegendentry{CouchDB};
        \addplot+[color=brown] [error bars/.cd, y dir = both, y explicit] table [x=NO_DOCS, y=COUCHBASE_AVG, y error = COUCHBASE_STD, col sep=comma] {q1_kw1.csv};
        \addlegendentry{Couchbase};
	\end{axis}
    \end{tikzpicture}
    \caption{$Q^{1}_{1}$}
    \label{fig:q1_kw1}   
    \end{subfigure}%
	\begin{subfigure}{0.5\columnwidth}
    	\begin{tikzpicture}[]
        \begin{axis}[
        xmin=0,
        xmax=5,
        ymin=0,
        ymax=450,
        bar width=2pt,
        ybar = 2pt,
        xtick \empty, 
        extra x ticks={0,1,2,3,4,5}, 
        extra x tick labels={$SF$,0.125,0.25,0.5,1},
        ylabel={Response time (s)},
		legend style={anchor=north, legend columns=3, legend cell align=left, at={(0.5,1.35)}},
        ]
        \addplot+[color=blue] [error bars/.cd, y dir = both, y explicit] table [x=NO_DOCS, y=BASEX_AVG, y error = BASEX_STD, col sep=comma] {q1_kw2.csv};
        \addlegendentry{BaseX};
        \addplot+[color=red] [error bars/.cd, y dir = both, y explicit] table [x=NO_DOCS, y=EXISTDB_AVG, y error = EXISTDB_STD, col sep=comma] {q1_kw2.csv};
        \addlegendentry{eXist-db};
        \addplot+[color=green] [error bars/.cd, y dir = both, y explicit] table [x=NO_DOCS, y=SEDNA_AVG, y error = SEDNA_STD, col sep=comma] {q1_kw2.csv};
        \addlegendentry{Sedna}
        \addplot+[color=black] [error bars/.cd, y dir = both, y explicit] table [x=NO_DOCS, y=MONGODB_AVG, y error = MONGODB_STD, col sep=comma] {q1_kw2.csv};
        \addlegendentry{MongoDB}; 
        \addplot+[color=pink] [error bars/.cd, y dir = both, y explicit] table [x=NO_DOCS, y=COUCHDB_AVG, y error = COUCHDB_STD, col sep=comma] {q1_kw2.csv};
        \addlegendentry{CouchDB};
        \addplot+[color=brown] [error bars/.cd, y dir = both, y explicit] table [x=NO_DOCS, y=COUCHBASE_AVG, y error = COUCHBASE_STD, col sep=comma] {q1_kw2.csv};
        \addlegendentry{Couchbase};
	\end{axis}
    \end{tikzpicture}
    \caption{$Q^{2}_{1}$}
    \label{fig:q1_kw2}   
    \end{subfigure}
    \begin{subfigure}{0.5\columnwidth}
   		\begin{tikzpicture}[]
        \begin{axis}[
        xmin=0,
        xmax=5,
        ymin=0,
        ymax=250,
        bar width=2pt,
        ybar = 2pt,
        xtick \empty, 
        extra x ticks={0,1,2,3,4,5}, 
        extra x tick labels={$SF$,0.125,0.25,0.5,1},
        ylabel={Response time (s)},
		legend style={anchor=north, legend columns=3, legend cell align=left, at={(0.5,1.35)}},
        ]
        \addplot+[color=blue] [error bars/.cd, y dir = both, y explicit] table [x=NO_DOCS, y=BASEX_AVG, y error = BASEX_STD, col sep=comma] {q1_kw3.csv};
        \addlegendentry{BaseX};
        \addplot+[color=red] [error bars/.cd, y dir = both, y explicit] table [x=NO_DOCS, y=EXISTDB_AVG, y error = EXISTDB_STD, col sep=comma] {q1_kw3.csv};
        \addlegendentry{eXist-db}; 
        \addplot+[color=green] [error bars/.cd, y dir = both, y explicit] table [x=NO_DOCS, y=SEDNA_AVG, y error = SEDNA_STD, col sep=comma] {q1_kw3.csv};
        \addlegendentry{Sedna}; 
        \addplot+[color=black] [error bars/.cd, y dir = both, y explicit] table [x=NO_DOCS, y=MONGODB_AVG, y error = MONGODB_STD, col sep=comma] {q1_kw3.csv};
        \addlegendentry{MongoDB}; 
        \addplot+[color=pink] [error bars/.cd, y dir = both, y explicit] table [x=NO_DOCS, y=COUCHDB_AVG, y error = COUCHDB_STD, col sep=comma] {q1_kw3.csv};
        \addlegendentry{CouchDB}; 
        \addplot+[color=brown] [error bars/.cd, y dir = both, y explicit] table [x=NO_DOCS, y=COUCHBASE_AVG, y error = COUCHBASE_STD, col sep=comma] {q1_kw3.csv};
        \addlegendentry{Couchbase};
	\end{axis}
	\end{tikzpicture}
    \caption{$Q^{3}_{1}$}
    \label{fig:q1_kw3}
    \end{subfigure}%
    \caption{Response time for $Q^{i}_{1}$}
    \label{fig:q1}
\end{figure*}
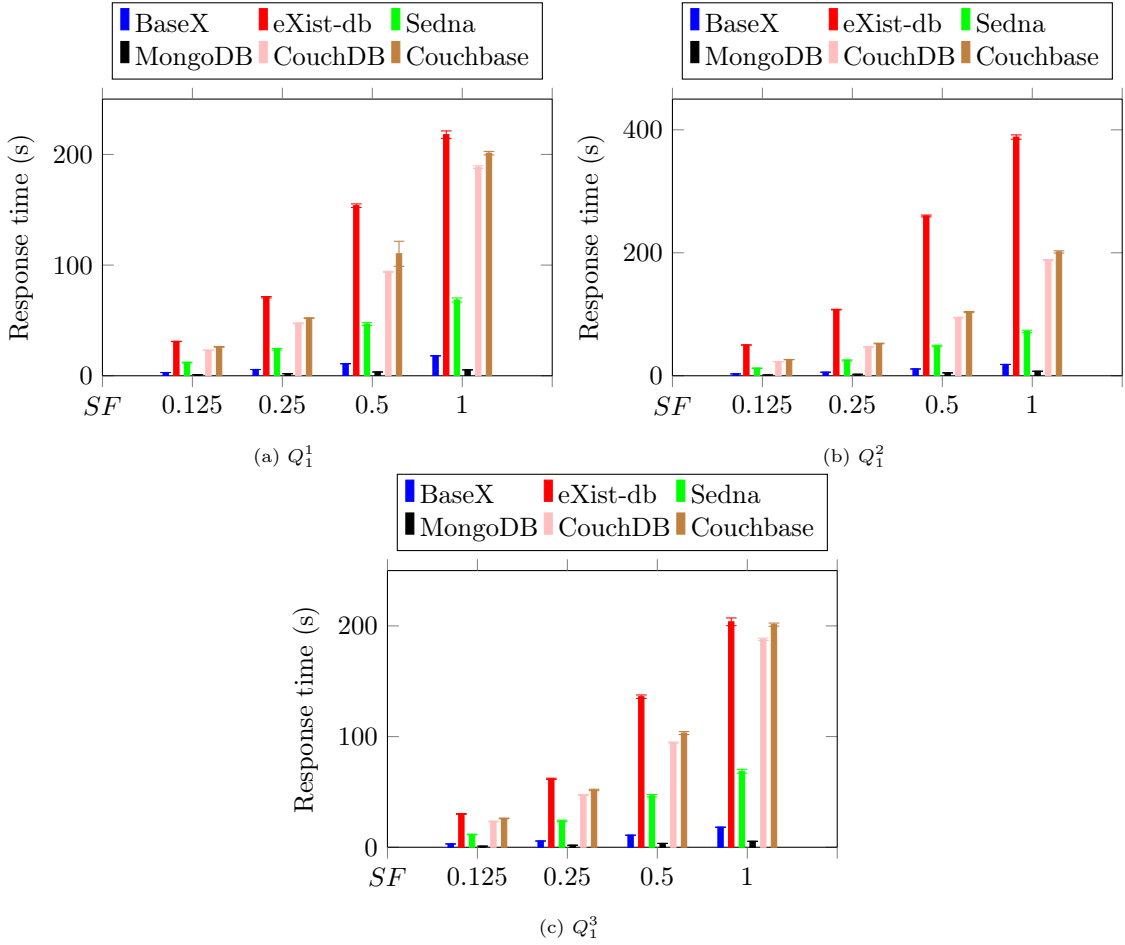

Figure~\ref{fig:q2q3} presents the results of $Q^{ij}_2$ and $Q^{ij}_3$ queries where $i$ and $j$ indicate the $t_{i}$ and $t_{j}$ keywords used for filtering (Table~\ref{tbl:params}) with $i = \overline{1,3}, j = \overline{1,3}$, and $i \neq j$.
For this set of queries, MongoDB has the best overall time performance regardless of the $SF$ factor.
BaseX achieves the second overall best performance and the best performance among the tested XDBMSes, regardless of the $SF$.
For the $Q^{ij}_{2}$ set of queries, the time performance of MongoDB has a factor between $\sim3.2$x and $\sim3.6$x faster then BaseX w.r.t. $SF$.
For the $Q^{ij}_{3}$ set of queries, the time performance of MongoDB has a factor between $\sim1.8$x and $\sim2.2$x faster then BaseX w.r.t. $SF$.

Couchbase presents the highest execution time for the $Q^{ij}_{2}$ queries regardless of $SF$, followed by the execution time of CouchDB.
CouchDB time performance is with a factor of $\sim1.2$x and $\sim1.1$x faster than Couchbase for the $Q^{ij}_{2}$, respectively $Q^{ij}_{3}$ queries regardless of $SF$.
The eXist-db XDBMS has the worst performance for the $Q^{ij}_{3}$ set of queries regardless of the $SF$.
For the $Q^{ij}_{2}$ set of queries, Sedna time performance has a factor of $\sim2$x better than CouchDB and a factor of $2$x worse than eXist-db.
For the $Q^{ij}_{3}$ set of queries, Sedna's query execution time is with a factor of $\sim1.5$x better than CouchDB and with a factor of $\sim5$x worst than BaseX.

\pgfplotsset{height=0.70\textwidth, width=1.00\columnwidth,
/pgfplots/ybar legend/.style={
        /pgfplots/legend image code/.code={
        \draw[##1,/tikz/.cd,bar width=3pt,yshift=-0.2em,bar shift=0pt]
                plot coordinates {(0cm,0.8em)};},
},
}
\begin{figure*}[!t]
	\centering
    \begin{subfigure}{0.5\columnwidth}
    \begin{tikzpicture}[]
        \begin{axis}[
        xmin=0,
        xmax=5,
        ymin=0,
        ymax=250,
        bar width=2pt,
        ybar = 2pt,
        xtick \empty, 
        extra x ticks={0,1,2,3,4,5}, 
        extra x tick labels={$SF$,0.125,0.25,0.5,1},
        ylabel={Response time (s)},
		legend style={anchor=north, legend columns=3, legend cell align=left, at={(0.5,1.35)}},
        ]
        \addplot+[color=blue] [error bars/.cd, y dir = both, y explicit] table [x=NO_DOCS, y=BASEX_AVG, y error = BASEX_STD, col sep=comma] {q2_kw1a2.csv};
        \addlegendentry{BaseX};
        \addplot+[color=red] [error bars/.cd, y dir = both, y explicit] table [x=NO_DOCS, y=EXISTDB_AVG, y error = EXISTDB_STD, col sep=comma] {q2_kw1a2.csv};
        \addlegendentry{eXist-db}; 
        \addplot+[color=green] [error bars/.cd, y dir = both, y explicit] table [x=NO_DOCS, y=SEDNA_AVG, y error = SEDNA_STD, col sep=comma] {q2_kw1a2.csv};
        \addlegendentry{Sedna}; 
        \addplot+[color=black] [error bars/.cd, y dir = both, y explicit] table [x=NO_DOCS, y=MONGODB_AVG, y error = MONGODB_STD, col sep=comma] {q2_kw1a2.csv};
        \addlegendentry{MongoDB}; 
        \addplot+[color=pink] [error bars/.cd, y dir = both, y explicit] table [x=NO_DOCS, y=COUCHDB_AVG, y error = COUCHDB_STD, col sep=comma] {q2_kw1a2.csv};
        \addlegendentry{CouchDB};
        \addplot+[color=brown] [error bars/.cd, y dir = both, y explicit] table [x=NO_DOCS, y=COUCHBASE_AVG, y error = COUCHBASE_STD, col sep=comma] {q2_kw1a2.csv};
        \addlegendentry{Couchbase};
        \end{axis}
        \end{tikzpicture}
        \caption{$Q^{12}_2 = Q^{1}_{1} \cap Q^{2}_{1}$}
    	\label{fig:q2_kw1a2}
    \end{subfigure}%
    \begin{subfigure}{0.5\columnwidth}
		\begin{tikzpicture}[]
        \begin{axis}[
        xmin=0,
        xmax=5,
        ymin=0,
        ymax=600,
        bar width=2pt,
        ybar = 2pt,
        xtick \empty, 
        extra x ticks={0,1,2,3,4,5}, 
        extra x tick labels={$SF$,0.125,0.25,0.5,1},
        ylabel={Response time (s)},
        legend style={anchor=north, legend columns=3, legend cell align=left, at={(0.5,1.35)}},
        ]
        \addplot+[color=blue] [error bars/.cd, y dir = both, y explicit] table [x=NO_DOCS, y=BASEX_AVG, y error = BASEX_STD, col sep=comma] {q2_kw1o2.csv};
        \addlegendentry{BaseX};
        \addplot+[color=red] [error bars/.cd, y dir = both, y explicit] table [x=NO_DOCS, y=EXISTDB_AVG, y error = EXISTDB_STD, col sep=comma] {q2_kw1o2.csv};
        \addlegendentry{eXist-db}; 
        \addplot+[color=green] [error bars/.cd, y dir = both, y explicit] table [x=NO_DOCS, y=SEDNA_AVG, y error = SEDNA_STD, col sep=comma] {q2_kw1o2.csv};
        \addlegendentry{Sedna}; 
         \addplot+[color=black] [error bars/.cd, y dir = both, y explicit] table [x=NO_DOCS, y=MONGODB_AVG, y error = MONGODB_STD, col sep=comma] {q2_kw1o2.csv};
        \addlegendentry{MongoDB}; 
         \addplot+[color=pink] [error bars/.cd, y dir = both, y explicit] table [x=NO_DOCS, y=COUCHDB_AVG, y error = COUCHDB_STD, col sep=comma] {q2_kw1o2.csv};
        \addlegendentry{CouchDB}; 
        \addplot+[color=brown] [error bars/.cd, y dir = both, y explicit] table [x=NO_DOCS, y=COUCHBASE_AVG, y error = COUCHBASE_STD, col sep=comma] {q2_kw1o2.csv};
        \addlegendentry{Couchbase};
        \end{axis}
        \end{tikzpicture}
        \caption{$ Q^{12}_{3} = Q^{1}_{1} \cup Q^{2}_{1}$}
    	\label{fig:q3_kw1o2}
    \end{subfigure}
    \begin{subfigure}{0.5\columnwidth}
    \begin{tikzpicture}[]
        \begin{axis}[
        xmin=0,
        xmax=5,
        ymin=0,
        ymax=250,
        bar width=2pt,
        ybar = 2pt,
        xtick \empty, 
        extra x ticks={0,1,2,3,4,5}, 
        extra x tick labels={$SF$,0.125,0.25,0.5,1},
        ylabel={Response time (s)},
		legend style={anchor=north, legend columns=3, legend cell align=left, at={(0.5,1.35)}},
        ]
        \addplot+[color=blue] [error bars/.cd, y dir = both, y explicit] table [x=NO_DOCS, y=BASEX_AVG, y error = BASEX_STD, col sep=comma] {q2_kw1a3.csv};
        \addlegendentry{BaseX};
        \addplot+[color=red] [error bars/.cd, y dir = both, y explicit] table [x=NO_DOCS, y=EXISTDB_AVG, y error = EXISTDB_STD, col sep=comma] {q2_kw1a3.csv};
        \addlegendentry{eXist-db}; 
        \addplot+[color=green] [error bars/.cd, y dir = both, y explicit] table [x=NO_DOCS, y=SEDNA_AVG, y error = SEDNA_STD, col sep=comma] {q2_kw1a3.csv};
        \addlegendentry{Sedna}; 
         \addplot+[color=black] [error bars/.cd, y dir = both, y explicit] table [x=NO_DOCS, y=MONGODB_AVG, y error = MONGODB_STD, col sep=comma] {q2_kw1a3.csv};
        \addlegendentry{MongoDB}; 
         \addplot+[color=pink] [error bars/.cd, y dir = both, y explicit] table [x=NO_DOCS, y=COUCHDB_AVG, y error = COUCHDB_STD, col sep=comma] {q2_kw1a3.csv};
        \addlegendentry{CouchDB}; 
        \addplot+[color=brown] [error bars/.cd, y dir = both, y explicit] table [x=NO_DOCS, y=COUCHBASE_AVG, y error = COUCHBASE_STD, col sep=comma] {q2_kw1a3.csv};
        \addlegendentry{Couchbase};
        \end{axis}
        \end{tikzpicture}
        \caption{$Q^{13}_{2} = Q^{1}_{1} \cap Q^{3}_{1}$}
    	\label{fig:q2_kw1a3}
    \end{subfigure}%
    \begin{subfigure}{0.5\columnwidth}
		\begin{tikzpicture}[]
        \begin{axis}[
        xmin=0,
        xmax=5,
        ymin=0,
        ymax=600,
        bar width=2pt,
        ybar = 2pt,
        xtick \empty, 
        extra x ticks={0,1,2,3,4,5}, 
        extra x tick labels={$SF$,0.125,0.25,0.5,1},
        ylabel={Response time (s)},
        legend style={anchor=north, legend columns=3, legend cell align=left, at={(0.5,1.35)}},
        ]
        \addplot+[color=blue] [error bars/.cd, y dir = both, y explicit] table [x=NO_DOCS, y=BASEX_AVG, y error = BASEX_STD, col sep=comma] {q2_kw1o3.csv};
        \addlegendentry{BaseX};
        \addplot+[color=red] [error bars/.cd, y dir = both, y explicit] table [x=NO_DOCS, y=EXISTDB_AVG, y error = EXISTDB_STD, col sep=comma] {q2_kw1o3.csv};
        \addlegendentry{eXist-db};
        \addplot+[color=green] [error bars/.cd, y dir = both, y explicit] table [x=NO_DOCS, y=SEDNA_AVG, y error = SEDNA_STD, col sep=comma] {q2_kw1o3.csv};
        \addlegendentry{Sedna}; 
         \addplot+[color=black] [error bars/.cd, y dir = both, y explicit] table [x=NO_DOCS, y=MONGODB_AVG, y error = MONGODB_STD, col sep=comma] {q2_kw1o3.csv};
        \addlegendentry{MongoDB}; 
         \addplot+[color=pink] [error bars/.cd, y dir = both, y explicit] table [x=NO_DOCS, y=COUCHDB_AVG, y error = COUCHDB_STD, col sep=comma] {q2_kw1o3.csv};
        \addlegendentry{CouchDB};
        \addplot+[color=brown] [error bars/.cd, y dir = both, y explicit] table [x=NO_DOCS, y=COUCHBASE_AVG, y error = COUCHBASE_STD, col sep=comma] {q2_kw1o3.csv};
        \addlegendentry{Couchbase};
        \end{axis}
        \end{tikzpicture}
        \caption{$Q^{13}_{3} = Q^{1}_{1} \cup Q^{3}_{1}$}
    	\label{fig:q3_kw1o3}
    \end{subfigure}
    \begin{subfigure}{0.5\columnwidth}
    \begin{tikzpicture}[]
        \begin{axis}[
        xmin=0,
        xmax=5,
        ymin=0,
        ymax=250,
        bar width=2pt,
        ybar = 2pt,
        xtick \empty, 
        extra x ticks={0,1,2,3,4,5}, 
        extra x tick labels={$SF$,0.125,0.25,0.5,1},
        ylabel={Response time (s)},
		legend style={anchor=north, legend columns=3, legend cell align=left, at={(0.5,1.35)}},
        ]
        \addplot+[color=blue] [error bars/.cd, y dir = both, y explicit] table [x=NO_DOCS, y=BASEX_AVG, y error = BASEX_STD, col sep=comma] {q2_kw2a3.csv};
        \addlegendentry{BaseX};
        \addplot+[color=red] [error bars/.cd, y dir = both, y explicit] table [x=NO_DOCS, y=EXISTDB_AVG, y error = EXISTDB_STD, col sep=comma] {q2_kw2a3.csv};
        \addlegendentry{eXist-db}; 
        \addplot+[color=green] [error bars/.cd, y dir = both, y explicit] table [x=NO_DOCS, y=SEDNA_AVG, y error = SEDNA_STD, col sep=comma] {q2_kw2a3.csv};
        \addlegendentry{Sedna}; 
         \addplot+[color=black] [error bars/.cd, y dir = both, y explicit] table [x=NO_DOCS, y=MONGODB_AVG, y error = MONGODB_STD, col sep=comma] {q2_kw2a3.csv};
        \addlegendentry{MongoDB}; 
         \addplot+[color=pink] [error bars/.cd, y dir = both, y explicit] table [x=NO_DOCS, y=COUCHDB_AVG, y error = COUCHDB_STD, col sep=comma] {q2_kw2a3.csv};
        \addlegendentry{CouchDB}; 
        \addplot+[color=brown] [error bars/.cd, y dir = both, y explicit] table [x=NO_DOCS, y=COUCHBASE_AVG, y error = COUCHBASE_STD, col sep=comma] {q2_kw2a3.csv};
        \addlegendentry{Couchbase};
        \end{axis}
        \end{tikzpicture}
        \caption{$Q^{23}_{2} = Q^{2}_{1} \cap Q^{3}_{1}$}
    	\label{fig:q2_kw2a3}
    \end{subfigure}%
    \begin{subfigure}{0.5\columnwidth}
		\begin{tikzpicture}[]
        \begin{axis}[
        xmin=0,
        xmax=5,
        ymin=0,
        ymax=600,
        bar width=2pt,
        ybar = 2pt,
        xtick \empty, 
        extra x ticks={0,1,2,3,4,5}, 
        extra x tick labels={$SF$,0.125,0.25,0.5,1},
        ylabel={Response time (s)},
        legend style={anchor=north, legend columns=3, legend cell align=left, at={(0.5,1.35)}},
        ]
        \addplot+[color=blue] [error bars/.cd, y dir = both, y explicit] table [x=NO_DOCS, y=BASEX_AVG, y error = BASEX_STD, col sep=comma] {q2_kw2o3.csv};
        \addlegendentry{BaseX};
        \addplot+[color=red] [error bars/.cd, y dir = both, y explicit] table [x=NO_DOCS, y=EXISTDB_AVG, y error = EXISTDB_STD, col sep=comma] {q2_kw2o3.csv};
        \addlegendentry{eXist-db}; 
        \addplot+[color=green] [error bars/.cd, y dir = both, y explicit] table [x=NO_DOCS, y=SEDNA_AVG, y error = SEDNA_STD, col sep=comma] {q2_kw2o3.csv};
        \addlegendentry{Sedna}; 
         \addplot+[color=black] [error bars/.cd, y dir = both, y explicit] table [x=NO_DOCS, y=MONGODB_AVG, y error = MONGODB_STD, col sep=comma] {q2_kw2o3.csv};
        \addlegendentry{MongoDB}; 
         \addplot+[color=pink] [error bars/.cd, y dir = both, y explicit] table [x=NO_DOCS, y=COUCHDB_AVG, y error = COUCHDB_STD, col sep=comma] {q2_kw2o3.csv};
        \addlegendentry{CouchDB}; 
        \addplot+[color=brown] [error bars/.cd, y dir = both, y explicit] table [x=NO_DOCS, y=COUCHBASE_AVG, y error = COUCHBASE_STD, col sep=comma] {q2_kw2o3.csv};
        \addlegendentry{Couchbase};
        \end{axis}
        \end{tikzpicture}
        \caption{$Q^{23}_{3} = Q^{2}_{1} \cup Q^{3}_{1}$}
    	\label{fig:q3_kw2o3}
    \end{subfigure}
    \caption{Response time for $Q^{ij}_{2}$ and $Q^{ij}_{3}$}
    \label{fig:q2q3}
\end{figure*}
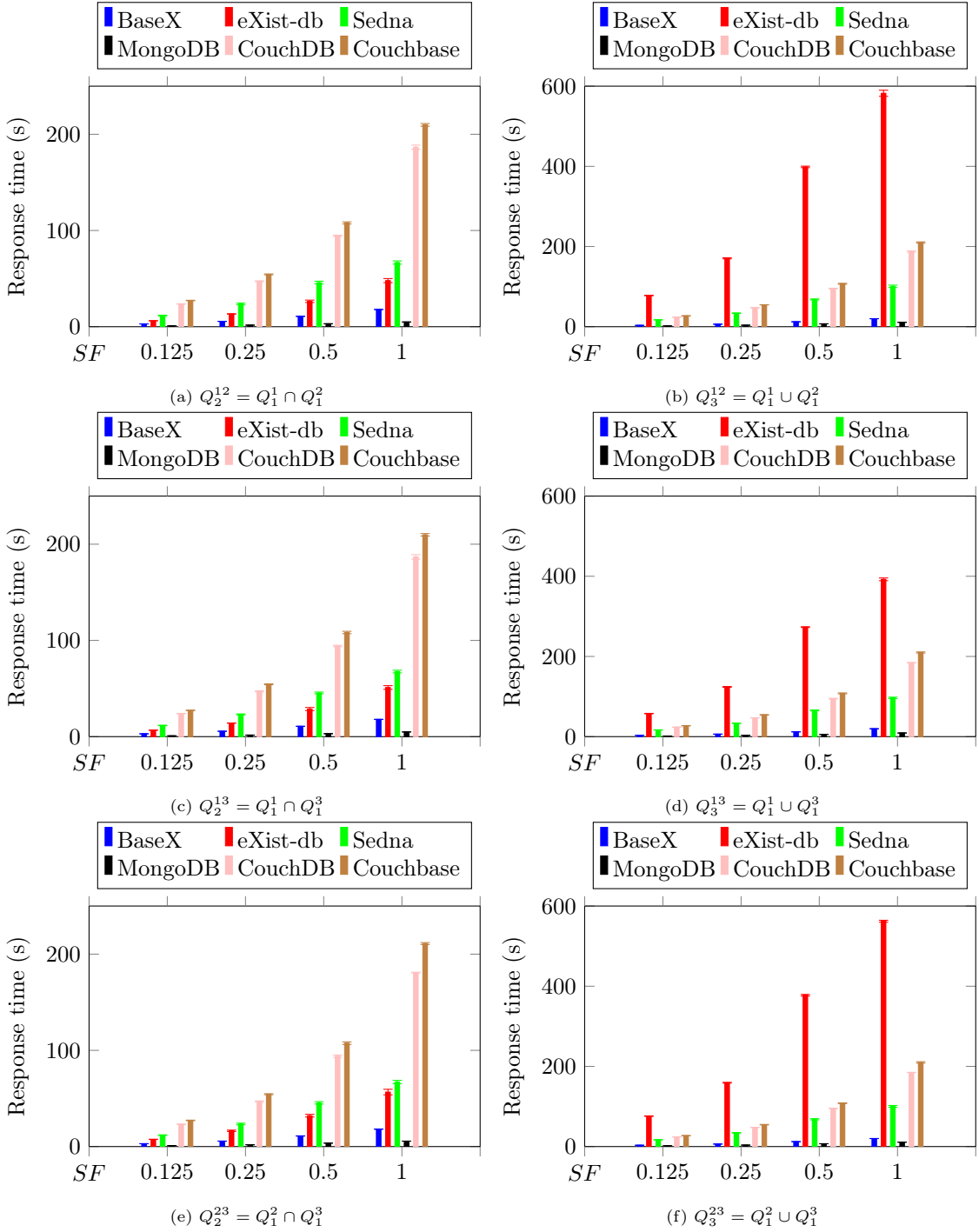 

Figure~\ref{fig:q4q5} presents the time performance of $Q_{4}$ and $Q_{5}$ queries for each DODBMS w.r.t. $SF$.
The time performance trend for $Q_{4}$ and $Q_{5}$ remains similar to the ones for $Q^{ij}_{2}$ and $Q^{ij}_{3}$, respectively.
CouchDB time performance is with a factor of $\sim1.3$x and $\sim1.2$x faster than Couchbase for the $Q^{ij}_{2}$, respectively $Q^{ij}_{3}$ queries regardless of $SF$.
MongoDB achieves the overall best time performance for both queries.
BaseX has the second-best time performance among the tested DODBMSes and the best performance among the XDBMSes.

\pgfplotsset{height=0.70\textwidth, width=1.00\columnwidth,
/pgfplots/ybar legend/.style={
        /pgfplots/legend image code/.code={
        \draw[##1,/tikz/.cd,bar width=3pt,yshift=-0.2em,bar shift=0pt]
                plot coordinates {(0cm,0.8em)};},
},
}
\begin{figure*}[!ht]
	\centering
    \begin{subfigure}{0.5\columnwidth}
    \begin{tikzpicture}[]
        \begin{axis}[
        xmin=0,
        xmax=5,
        ymin=0,
        ymax=250,
        bar width=2pt,
        ybar = 2pt,
        xtick \empty, 
        extra x ticks={0,1,2,3,4,5}, 
        extra x tick labels={$SF$,0.125,0.25,0.5,1},
        ylabel={Response time (s)},
		legend style={anchor=north, legend columns=3, legend cell align=left, at={(0.5,1.35)}},
        ]
        \addplot+[color=blue] [error bars/.cd, y dir = both, y explicit] table [x=NO_DOCS, y=BASEX_AVG, y error = BASEX_STD, col sep=comma] {q3_kw1a2a3.csv};
        \addlegendentry{BaseX};
        \addplot+[color=red] [error bars/.cd, y dir = both, y explicit] table [x=NO_DOCS, y=EXISTDB_AVG, y error = EXISTDB_STD, col sep=comma] {q3_kw1a2a3.csv};
        \addlegendentry{eXist-db}; 
        \addplot+[color=green] [error bars/.cd, y dir = both, y explicit] table [x=NO_DOCS, y=SEDNA_AVG, y error = SEDNA_STD, col sep=comma] {q3_kw1a2a3.csv};
        \addlegendentry{Sedna}; 
        \addplot+[color=black] [error bars/.cd, y dir = both, y explicit] table [x=NO_DOCS, y=MONGODB_AVG, y error = MONGODB_STD, col sep=comma] {q3_kw1a2a3.csv};
        \addlegendentry{MongoDB}; 
        \addplot+[color=pink] [error bars/.cd, y dir = both, y explicit] table [x=NO_DOCS, y=COUCHDB_AVG, y error = COUCHDB_STD, col sep=comma] {q3_kw1a2a3.csv};
        \addlegendentry{CouchDB}; 
        \addplot+[color=brown] [error bars/.cd, y dir = both, y explicit] table [x=NO_DOCS, y=COUCHBASE_AVG, y error = COUCHBASE_STD, col sep=comma] {q3_kw1a2a3.csv};
        \addlegendentry{Couchbase};
        \end{axis}
        \end{tikzpicture}
        \caption{$Q_{4} = Q^{1}_{1} \cap Q^{2}_{1} \cap Q^{3}_{1}$}
    	\label{fig:q4}
    \end{subfigure}%
    \begin{subfigure}{0.5\columnwidth}
		\begin{tikzpicture}[]
        \begin{axis}[
        xmin=0,
        xmax=5,
        ymin=0,
        ymax=800,
        bar width=2pt,
        ybar = 2pt,
        xtick \empty, 
        extra x ticks={0,1,2,3,4,5}, 
        extra x tick labels={$SF$,0.125,0.25,0.5,1},
        ylabel={Response time (s)},
        legend style={anchor=north, legend columns=3, legend cell align=left, at={(0.5,1.35)}},
        ]
        \addplot+[color=blue] [error bars/.cd, y dir = both, y explicit] table [x=NO_DOCS, y=BASEX_AVG, y error = BASEX_STD, col sep=comma] {q3_kw1o2o3.csv};
        \addlegendentry{BaseX};
        \addplot+[color=red] [error bars/.cd, y dir = both, y explicit] table [x=NO_DOCS, y=EXISTDB_AVG, y error = EXISTDB_STD, col sep=comma] {q3_kw1o2o3.csv};
        \addlegendentry{eXist-db}; 
        \addplot+[color=green] [error bars/.cd, y dir = both, y explicit] table [x=NO_DOCS, y=SEDNA_AVG, y error = SEDNA_STD, col sep=comma] {q3_kw1o2o3.csv};
        \addlegendentry{Sedna}; 
         \addplot+[color=black] [error bars/.cd, y dir = both, y explicit] table [x=NO_DOCS, y=MONGODB_AVG, y error = MONGODB_STD, col sep=comma] {q3_kw1o2o3.csv};
        \addlegendentry{MongoDB}; 
        \addplot+[color=pink] [error bars/.cd, y dir = both, y explicit] table [x=NO_DOCS, y=COUCHDB_AVG, y error = COUCHDB_STD, col sep=comma] {q3_kw1o2o3.csv};
        \addlegendentry{CouchDB}; 
        \addplot+[color=brown] [error bars/.cd, y dir = both, y explicit] table [x=NO_DOCS, y=COUCHBASE_AVG, y error = COUCHBASE_STD, col sep=comma] {q3_kw1o2o3.csv};
        \addlegendentry{Couchbase};
        \end{axis}
        \end{tikzpicture}
        \caption{$Q_{5} = Q^{1}_{1} \cup Q^{2}_{1}) \cup Q^{3}_{1}$}
    	\label{fig:q5}
    \end{subfigure}
    \caption{Response time for $Q_{4}$ and $Q_{5}$}
    \label{fig:q4q5}
\end{figure*}
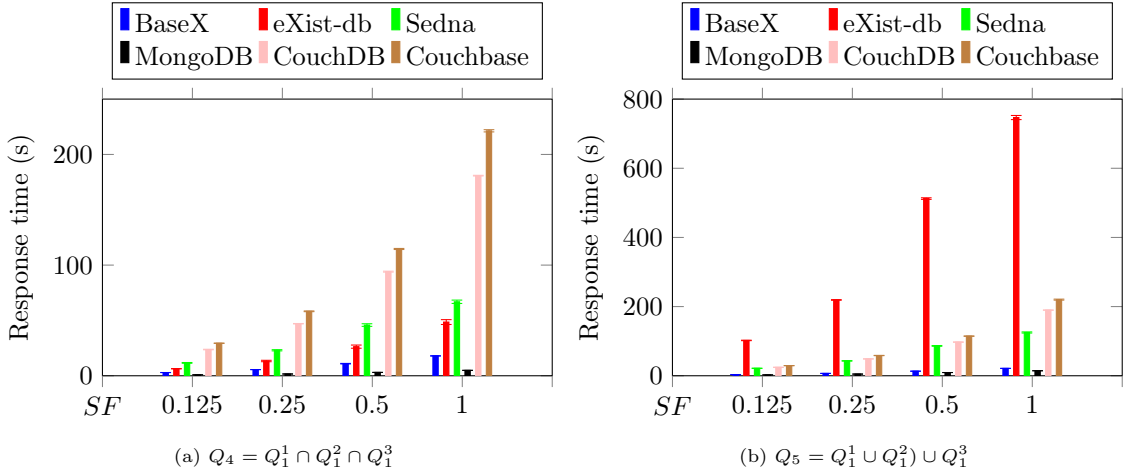

Figure~\ref{fig:q_aggreg} shows the results for the aggregation queries, i.e., $Q_{6}$ to $Q_{9}$.
For the queries $Q_{6}$, $Q_{7}$, and $Q_9$, BaseX has the best time performance and significantly outperforms MongoDB and CouchDB with a factor of $\sim2$x, regardless of the $SF$.
For the $Q_{8}$ query, CouchDB achieved the best query execution time, while Couchbase the worst.
MongoDB has the second best query response time among the studied DODBMSes for $Q_{6}$, $Q_{7}$, and $Q_9$.
MongoDB's response time for these queries is almost on parity with the response time of CouchDB w.r.t. $SF$, although MongoDB executes the aggregation functions at runtime.

For $Q_{7}$, Couchbase has a large standard deviation.
During testing, this query finished with the error "Index scan timed out".
The tests that finished with the status "success" returned fluctuating time performance for each run.
This abnormal behavior of the Couchbase system can be sometimes observed for complex queries on large collections.

For $Q_{8}$ which has the highest selectivity, CouchDB holds the best time performance.
We attribute this result to the mechanism used by CouchDB to store aggregation functions.
Aggregation functions are stored in materialized views also named indices in CouchDB.
Using this technique, CouchDB manages to outperform BaseX and MongoDB, which execute aggregation functions at runtime, for queries with high selectivity.
With Couchbase, the complexity and selectivity together with the \texttt{UNNEST} clause required to extract the nested documents in order to filter and group the information, increases the runtime significantly while decreasing the overall query performance.

The aggregation queries did not work on Sedna.
When executing these queries, the XDBMS remained unresponsive for days, and we had to manually stop the system, the related services, and the background processes.
We note that Sedna also executes aggregation functions at runtime.
\textsc{We suspect that one reason for Sedna's failure to execute the aggregation queries is also the outdated XQuery 1.0 query language.}

The eXist-db XDBMS has the highest query time for $Q_{6}$, $Q_{7}$, and $Q_9$ queries.
The execution is done at runtime.
For this XDBMS, query $Q_{7}$ worked only for $SF=0.125$.
For other $SF$ values, the query returned memory errors, although we have tuned this XDBMS to work with the same parameters as the other DODBMSes.
Thus, eXist-db is highly dependent on the JVM (Java Virtual Machine) memory allocation mechanism.

\pgfplotsset{height=0.70\textwidth, width=1.00\columnwidth,
/pgfplots/ybar legend/.style={
        /pgfplots/legend image code/.code={
        \draw[##1,/tikz/.cd,bar width=3pt,yshift=-0.2em,bar shift=0pt]
                plot coordinates {(0cm,0.8em)};},
},
}
\begin{figure*}[!ht]
	\centering
    \begin{subfigure}{0.5\columnwidth}
    \begin{tikzpicture}[]
        \begin{axis}[
        xmin=0,
        xmax=5,
        ymin=0,
        ymax=6000,
        bar width=2pt,
        ybar = 2pt,
        xtick \empty, 
        extra x ticks={0,1,2,3,4,5}, 
        extra x tick labels={$SF$,0.125,0.25,0.5,1},
        ylabel={Response time (s)},
		legend style={anchor=north, legend columns=3, legend cell align=left, at={(0.5,1.35)}},
        ]
        \addplot+[color=blue] [error bars/.cd, y dir = both, y explicit] table [x=NO_DOCS, y=BASEX_AVG, y error = BASEX_STD, col sep=comma] {count_docs_authors.csv};
        \addlegendentry{BaseX};
        \addplot+[color=red] [error bars/.cd, y dir = both, y explicit] table [x=NO_DOCS, y=EXISTDB_AVG, y error = EXISTDB_STD, col sep=comma] {count_docs_authors.csv};
        \addlegendentry{eXist-db}; 
        \addplot+[color=black] [error bars/.cd, y dir = both, y explicit] table [x=NO_DOCS, y=MONGODB_AVG, y error = MONGODB_STD, col sep=comma] {count_docs_authors.csv};
        \addlegendentry{MongoDB}; 
        \addplot+[color=pink] [error bars/.cd, y dir = both, y explicit] table [x=NO_DOCS, y=COUCHDB_AVG, y error = COUCHDB_STD, col sep=comma] {count_docs_authors.csv};
        \addlegendentry{ChouchDB}; 
        \addplot+[color=brown] [error bars/.cd, y dir = both, y explicit] table [x=NO_DOCS, y=COUCHBASE_AVG, y error = COUCHBASE_STD, col sep=comma] {count_docs_authors.csv};
        \addlegendentry{Couchbase};
        \end{axis}
        \end{tikzpicture}
        \caption{$Q_{6}$}
    	\label{fig:q6}
    \end{subfigure}%
    \begin{subfigure}{0.5\columnwidth}
		\begin{tikzpicture}[]
        \begin{axis}[
        xmin=0,
        xmax=5,
        ymin=0,
        ymax=4000,
        bar width=2pt,
        ybar = 2pt,
        xtick \empty, 
        extra x ticks={0,1,2,3,4,5}, 
        extra x tick labels={$SF$,0.125,0.25,0.5,1},
        ylabel={Response time (s)},
        legend style={anchor=north, legend columns=3, legend cell align=left, at={(0.5,1.35)}},
        ]
        \addplot+[color=blue] [error bars/.cd, y dir = both, y explicit] table [x=NO_DOCS, y=BASEX_AVG, y error = BASEX_STD, col sep=comma] {count_docs_authors_year.csv};
        \addlegendentry{BaseX};
        \addplot+[color=red] [error bars/.cd, y dir = both, y explicit] coordinates { (1, 3759.40) +- (0, 8.85) };
        \addlegendentry{eXist-db}; 
        \addplot+[color=black] [error bars/.cd, y dir = both, y explicit] table [x=NO_DOCS, y=MONGODB_AVG, y error = MONGODB_STD, col sep=comma] {count_docs_authors_year.csv};
        \addlegendentry{MongoDB}; 
        \addplot+[color=pink] [error bars/.cd, y dir = both, y explicit] table [x=NO_DOCS, y=COUCHDB_AVG, y error = COUCHDB_STD, col sep=comma] {count_docs_authors_year.csv};
        \addlegendentry{CouchDB};
        \addplot+[color=brown] [error bars/.cd, y dir = both, y explicit] table [x=NO_DOCS, y=COUCHBASE_AVG, y error = COUCHBASE_STD, col sep=comma] {count_docs_authors_year.csv};
        \addlegendentry{Couchbase};
        \end{axis}
        \end{tikzpicture}
        \caption{$Q_{7}$}
    	\label{fig:q7}
    \end{subfigure}
    \begin{subfigure}{0.5\columnwidth}
		\begin{tikzpicture}[]
        \begin{axis}[
        xmin=0,
        xmax=5,
        ymin=0,
        ymax=500,
        bar width=2pt,
        ybar = 2pt,
        xtick \empty, 
        extra x ticks={0,1,2,3,4,5}, 
        extra x tick labels={$SF$,0.125,0.25,0.5,1},
        ylabel={Response time (s)},
        legend style={anchor=north, legend columns=3, legend cell align=left, at={(0.5,1.35)}},
        ]
        \addplot+[color=blue] [error bars/.cd, y dir = both, y explicit] table [x=NO_DOCS, y=BASEX_AVG, y error = BASEX_STD, col sep=comma] {count_all_authors_year_kw1a2a3.csv};
        \addlegendentry{BaseX};
        \addplot+[color=red] [error bars/.cd, y dir = both, y explicit] table [x=NO_DOCS, y=EXISTDB_AVG, y error = EXISTDB_STD, col sep=comma] {count_all_authors_year_kw1a2a3.csv};
        \addlegendentry{eXist-db}; 
         \addplot+[color=black] [error bars/.cd, y dir = both, y explicit] table [x=NO_DOCS, y=MONGODB_AVG, y error = MONGODB_STD, col sep=comma] {count_all_authors_year_kw1a2a3.csv};
        \addlegendentry{MongoDB}; 
        \addplot+[color=pink] [error bars/.cd, y dir = both, y explicit] table [x=NO_DOCS, y=COUCHDB_AVG, y error = COUCHDB_STD, col sep=comma] {count_all_authors_year_kw1a2a3.csv};
        \addlegendentry{CouchDB}; 
        \addplot+[color=brown] [error bars/.cd, y dir = both, y explicit] table [x=NO_DOCS, y=COUCHBASE_AVG, y error = COUCHBASE_STD, col sep=comma] {count_all_authors_year_kw1a2a3.csv};
        \addlegendentry{Couchbase};
        \end{axis}
        \end{tikzpicture}
        \caption{$Q_{8}$}
    	\label{fig:q8}
    \end{subfigure}%
    \begin{subfigure}{0.5\columnwidth}
		\begin{tikzpicture}[]
        \begin{axis}[
        xmin=0,
        xmax=5,
        ymin=0,
        ymax=1200,
        bar width=2pt,
        ybar = 2pt,
        xtick \empty, 
        extra x ticks={0,1,2,3,4,5}, 
        extra x tick labels={$SF$,0.125,0.25,0.5,1},
        ylabel={Response time (s)},
        legend style={anchor=north, legend columns=3, legend cell align=left, at={(0.5,1.35)}},
        ]
        \addplot+[color=blue] [error bars/.cd, y dir = both, y explicit] table [x=NO_DOCS, y=BASEX_AVG, y error = BASEX_STD, col sep=comma] {count_all_authors_year_kw1o2o3.csv};
        \addlegendentry{BaseX};
        \addplot+[color=red] [error bars/.cd, y dir = both, y explicit] table [x=NO_DOCS, y=EXISTDB_AVG, y error = EXISTDB_STD, col sep=comma] {count_all_authors_year_kw1o2o3.csv};
        \addlegendentry{eXist-db}; 
        \addplot+[color=black] [error bars/.cd, y dir = both, y explicit] table [x=NO_DOCS, y=MONGODB_AVG, y error = MONGODB_STD, col sep=comma] {count_all_authors_year_kw1o2o3.csv};
        \addlegendentry{MongoDB}; 
        \addplot+[color=pink] [error bars/.cd, y dir = both, y explicit] table [x=NO_DOCS, y=COUCHDB_AVG, y error = COUCHDB_STD, col sep=comma] {count_all_authors_year_kw1o2o3.csv};
        \addlegendentry{CouchDB};
        \addplot+[color=brown] [error bars/.cd, y dir = both, y explicit] table [x=NO_DOCS, y=COUCHBASE_AVG, y error = COUCHBASE_STD, col sep=comma] {count_all_authors_year_kw1o2o3.csv};
        \addlegendentry{Couchbase};
        \end{axis}
        \end{tikzpicture}
        \caption{$Q_{9}$}
    	\label{fig:q9}
    \end{subfigure}
    \caption{Response time for aggregation queries}
    \label{fig:q_aggreg}
\end{figure*}
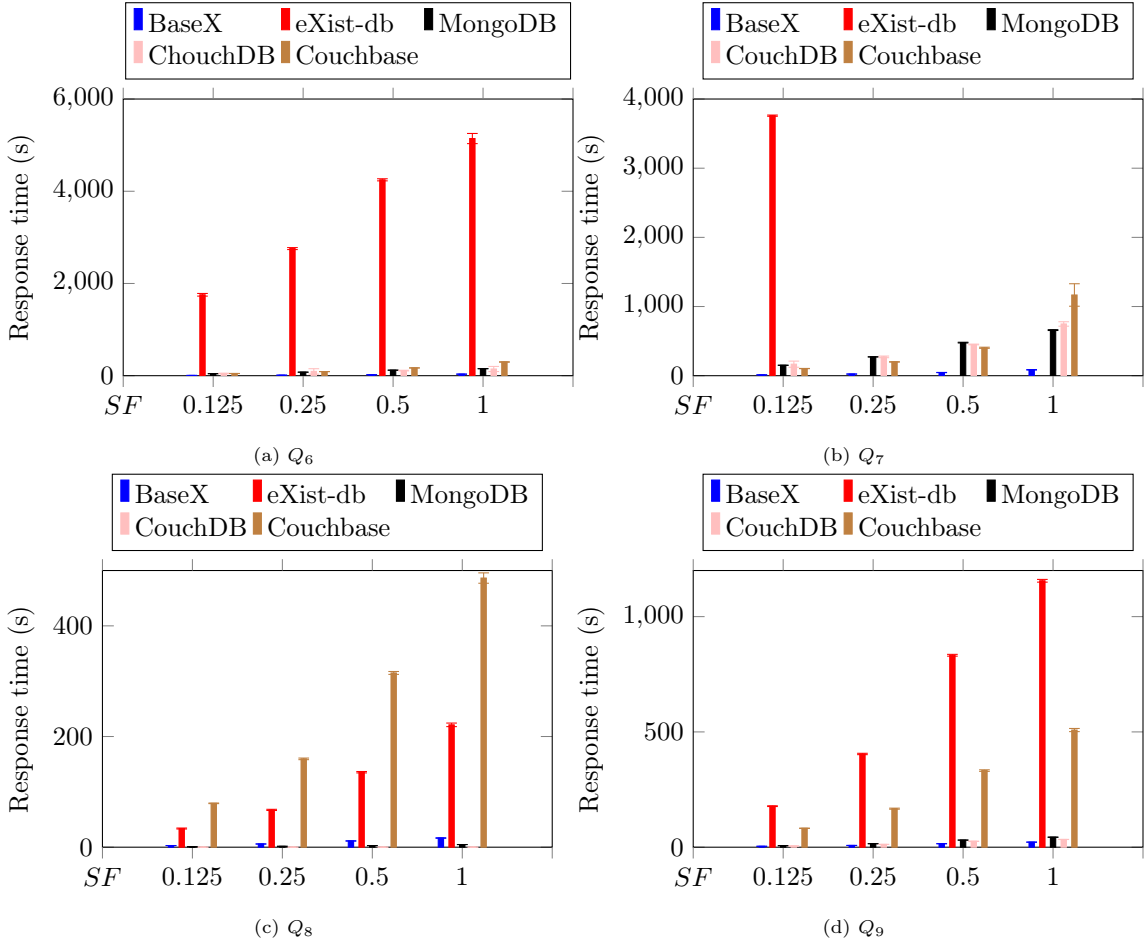 

\subsection{Discussions on the Experimental Design Choices}

In this study, we present our findings regarding the performance of filtering and aggregation queries on a large dataset for XDBMSes and JDBMSes w.r.t. different scale factors.
We observe that the XDBMSes perform as well as JDBMSes for specific use cases, with BaseX even outperforming the more popular JDBMSes on three out of the four aggregation queries.
Among the JDBMSes, MongoDB has the overall best performance.

For our comparison, we do not take into account horizontal scalability through sharding and replication, as not all of the analyzed DBMSes have such a functionality. 
Furthermore, it is essential first to understand single-node performance before considering horizontal scaling.
Thus, the aim of the paper is to examine single instance deployments.

There are many real-world scenarios where such single-instance deployment is preferred.
As a first example, XDBMSes can be used for fast application development, analyzing and querying log data, or storing and retrieving IoT sensor data.
XDBMSes are good candidates for storing large documents, managing long-running transactions, and querying hierarchical data structures in environments that require rapidly evolving schemes.
Furthermore, these DBMSes are lightweight and do not require dedicated hardware, software, or a lot of resources.
Thus, managing to lower resource costs at the data center site and enabling on-site data analysis and decision making.
Therefore, they can be utilized in Edge and Fog Computing with ease.

The creation of network islands due to faulty nodes is very common in the Edge/Fog environment.
Even in the presence of well-defined recovery mechanisms, the formation of temporal network islands is unfavorable for sharding, as the overall latency increases if nodes go down and then up again.
Hence, single-instance deployments are favored in these environments.

Another real-world scenario where such single-instance deployment can be used is for small to medium scale document management systems.
These management systems are useful to smaller enterprises, where data is kept in the company due to GDPR (European Union Legislation on General Data Protection Regulation).
Moreover, as in many cases most of the data is in semi-structured formats, such as XML and JSON, single instance DODBMSes are a good candidate for storing and managing such documents.
Thus, removing from the company’s costs the maintenance of a data center.

It is also important to mention that the focus of our benchmark is on data retrieval and not on write operations because, in real-world applications, multiple techniques can be put in check to balance the write operations and minimize the workload.
Moreover, data persistence can be achieved much later within a DBMS, depending on the workload and the systems write and logging mechanisms.

Furthermore, we loaded the data into the database using different methods.
Because not all of the tested DODBMSes have their own data load tools, we developed our own data loading programs.
By utilizing our data load programs and not native load DBMS functionalities, we added a new layer of complexity which decreases write performance.
This makes the loading process to be dependent on external DBC (database connectors) implementations, and not on the DODBMS internal functionalities.

\section{Conclusion}\label{sec:conclusion}

In this paper, we present an overview and comparison of DODBMSes that encode information using XML and JSON formats and propose a benchmark using filtering and aggregation queries on a heterogeneous dataset.
For our experiments we chose three XDBMSes, i.e., BaseX, eXist-db, Sedna, and three JDBMSes, i.e., MongoDB, CouchDB, and Couchbase.
These DODBMSes are open-source and free to use systems, whose license does not forbid benchmarking.

Our comparison focuses on key functionalities required by Big Data and IoT systems for storing and extracting information from large volumes of data.
For this comparison, we also consider the transactions' properties of each DODBMSes, their in-memory capabilities, and how these systems deal with atomicity, consistency, isolation, durability with regards to operations such as accessing, modifying, and saving documents.
We also present for each DODBMS its support for replication and partitioning of data and how it manages these Big Data requirements.
Furthermore, we present the querying languages used for extracting information as well as the different types of indices provided by each DODBMS to improve retrieval response time.

The proposed benchmark uses different queries to emphasize the time performance of DODBMSes and highlights the capabilities of XDBMSes and JDBMSes.
Furthermore, our solution proves its portability, scalability, and relevance by its design.
The benchmark is portable, as it works on multiple systems.
For this purpose, we compare the performance of several DODBMSes, i.e., BaseX, eXist-db, Sedna, MongoDB, CouchDB, and Couchbase.
To demonstrate the scalability of our solution, we introduced $SF$, the scaling factor that generates an incremental growth in the data volume for our experiments.
By increasing the queries' complexity together with the $SF$ factor, we analyze the behavior of the systems from the scaling perspective.
We observe that all the DODBMSes have a linear increase at runtime.
Furthermore, BaseX proves to be a good choice when dealing with aggregations.
Finally, our experimental results show that our benchmark is indeed relevant in comparing the runtime performance of different DODBMSes.

The performance tests provide some interesting and unexpected results.
Among the XDBMSes, BaseX has the best overall performance.
BaseX even outperforms the JDBMSes selected for this benchmark, i.e., MongoDB, CouchDB, and Couchbase, for three out of the four aggregation queries proposed.
We observe that Couchbase has the overall worst performance among the JDBMSes.
Sedna outperforms CouchDB and Couchbase when dealing with filtering queries, but does not work for the aggregation queries.
MongoDB has the overall best time performance for the filtering queries and it outperforms BaseX only for the aggregation query $Q_{8}$.
eXist-db has some strange behavior when dealing with both filtering and aggregation queries.
Also, it is highly dependent on the JVM, which needs to be tuned for each query, making this XDBMS hard to work with.
However, we can assume that eXist-db works well on a query to query basis.

Following the results obtained by the benchmark, we can answer the three research questions and conclude that XDBMSes are still useful: their performance is as good as JDBMSes and they are good candidates for Big Data Management.
Furthermore, XDBMSes are well-suited for several current real-world scenarios.
Firstly, XDBMSes are reliable systems for storing large documents, managing long-running transactions, and querying hierarchical data structures in Edge/Fog environments (e.g., smart agriculture, healthcare wearables, etc.), as these types of DODBMSes are lightweight and do not require dedicated hardware, software, or a lot of resources.
Secondly, XDBMSes can be used as small to medium scale document management systems in smaller enterprises, where data are kept in the company due to GDPR.
Thirdly, in the case of Big Data analysis, they prove to be well-suited when the documents are in XML format, by removing the ETL (Extract, Transform, Load) processes from the storing, managing, and analysis pipeline.

%As future work, we plan to improve the performance of our solution by designing new sampling strategies and aggregation queries. The sampling methods will include constraints on other labels and values contained in the records. Also, we aim to add more dimension for grouping, to boost the performance by lowering the query selectivity.

As future work, we plan to improve the support for OLAP queries~\cite{X1}  on XML data and XML data in combination with other data~\cite{X2,X3} both in terms of performance and functionality.
This includes designing new sampling 
strategies and supporting more aggregation queries~\cite{X2}. The sampling methods will include constraints on other labels
and values contained in the records. Also, we aim to add more dimension for grouping~\cite{X2}, to boost
the performance by lowering the query selectivity and performing query rewriting~\cite{X3}, and to add further grouping functionality~\cite{X2}.

% \section*{References}

\section*{Acknowledgement}

The research presented in this paper was supported in part by \textit{the Danish Independent Research Council}, through the \textit{SEMIOTIC} project, and the \textit{Robots and Society: Cognitive Systems for Personal Robots and Autonomous Vehicles (ROBIN)} project \textit{CCCDI-UEFISCDI grant No. PN-III-P1-1.2-PCCDI-2017-0734}.

\bibliographystyle{elsarticle-num}
\bibliography{bdr2021tadp}

\end{document}